\documentclass[twocolumn,superscriptaddress,floatfix,
longbibliography,aps,prb,preprintnumbers,10pt,groupedaddress]{revtex4-2}
\usepackage{titlesec}

\usepackage{graphicx}
\usepackage{amssymb}
\usepackage{bm}
\usepackage{graphicx,color}
\usepackage[urlcolor=blue,colorlinks=true,citecolor=blue,linkcolor=blue,pdfstartview={FitH},bookmarks=false]{hyperref}
\begin{document}

\title{Dimensional reduction of the Luttinger-Ward functional for spin-degenerate $D$-dimensional electron gases} 
	
\date{\today}
\author{Dmitry Miserev,$^{1\ast}$ Jelena Klinovaja,$^{1}$ and Daniel Loss$^{1}$}
\affiliation{$^{1}$Department of Physics, University of Basel, \\
Klingelbergstrasse 82, CH-4056 Basel, Switzerland\\
	}

\begin{abstract}
	We consider an isotropic spin-degenerate interacting uniform $D$-dimensional electron gas (DDEG) with $D > 1$ within the Luttinger-Ward (LW) formalism.
	We derive the asymptotically exact semiclassical/infrared limit of the LW functional at large distances, $r \gg \lambda_F$, and large times, $\tau \gg 1/E_F$, where $\lambda_F$ and $E_F$ are the Fermi wavelength and the Fermi energy, respectively.
	The LW functional is represented by skeleton diagrams, each skeleton diagram consists of appropriately connected dressed fermion loops.
	First, we prove that every $D$-dimensional skeleton diagram consisting of a single fermion loop is reduced to a one-dimensional (1D) fermion loop with the same diagrammatic structure, which justifies the name dimensional reduction.
	This statement, combined with the fermion loop cancellation theorem (FLCT), agrees with results of multidimensional bosonization.
	Here we show that the backscattering and the spectral curvature, both explicitly violate the FLCT and both are irrelevant for a 1DEG, become relevant at $D > 1$ and $D > 2$, respectively.
	The reason for this is a strong infrared divergence of the skeleton diagrams containing multiple fermion loops at $D > 1$.
	These diagrams, which are omitted within the multidimensional bosonization approaches, account for the non-collinear scattering processes.
	Thus, the dimensional reduction provides the framework to go
	beyond predictions of the multidimensional bosonization.
	A simple diagrammatic structure of the reduced LW functional is another advantage of our approach.
	The dimensional reduction technique is also applicable to the thermodynamic potential and various approximations, from perturbation theory to self-consistent approaches.
\end{abstract}

\maketitle
	
\section{Introduction}

Rigorous theoretical description of interacting electrons is an extreme theoretical challenge fostering the development of new approaches in interacting quantum field theories.
Exactly solvable interacting models provide a valuable insight into possible structure of strongly correlated electron matter, yet they normally rely on assumptions atypical of the realistic physical systems.
Among the most celebrated examples, there are various exactly solvable models with large number $N \gg 1$ of the electron 	flavors~\cite{ye,parcollet,sachdev,chowdhury,georges,abanov} or large number $D \gg 1$ of spatial dimensions~\cite{georgesrev,schiller,hoshino,janis}.
Another direction of theoretical research is to extend the one-dimensional (1D) bosonization technique~\cite{tomonaga,luttliq,lieb,dzyalo,haldanelut,giamarchi} to higher dimensions.
One way to do this is via so-called weakly coupled wire constructions~\cite{wen1990,teo,laubscher} where a $D$-dimensional electron system is represented as an array of weakly coupled Luttinger liquids.
The problem of this approach is that the electron hopping is strong only along the wire direction and must be treated as a small perturbation along other directions~\cite{wen1990}.
Coherent zero sound modes in Fermi liquids~\cite{mermin} provide a solid basis for the bosonization of a $D$-dimensional electron gas (DDEG)~\cite{luther,haldane,fradkin,neto,delacretaz}, this approach is also known as the Fermi surface (FS) bosonization.
The multidimensional bosonization was equivalently formulated within the functional integral approach \cite{frohlich,marchetti,schwiete,aleiner,efetov,pepin,meier}
and via the Ward's identity~\cite{castellani,kopietz,kopietzbook,metzner,houghton}, following the recipe of Ref.~\cite{dzyalo} for the 1DEG.
Explicit multidimensional bosonization solutions were found for the case of linear fermion spectrum near the FS and for forward-scattering interaction \cite{luther,haldane,fradkin,neto,delacretaz,frohlich,marchetti,schwiete,aleiner,efetov,pepin,meier,castellani,kopietz,kopietzbook,metzner,houghton},
these results are supported by the fermion loop cancellation theorem (FLCT) that is approximately valid under the aforementioned conditions~\cite{kopietz,neumayr,kopper}.
However, the backscattering is known to cause
infrared non-analyticities in the thermodynamic potential \cite{schwiete,aleiner,efetov,pepin,meier,maslov,zak1,zak2,miserev22}, indicating its importance for the low-energy physics in a DDEG.
The electron spectral curvature is shown to be qualitatively important in both 1DEG \cite{pustilnik,imambekov,schmidt} and 2DEG \cite{gangadharaiah,glazman} for the semiclassical/infrared asymptotics of some correlation functions and the low-temperature transport properties.
Here we consider both the forward- and the backscattering interactions as well as a general electron dispersion, which allows us to go beyond the FLCT which is no longer valid  if  backscattering and/or electron spectral curvature are present.

In this work we propose a new powerful theoretical tool, the \textit{dimensional	reduction procedure}, that complements the existing multidimensional bosonization approaches \cite{luther,haldane,fradkin,neto,delacretaz,frohlich,marchetti,schwiete,aleiner,efetov,pepin,meier,castellani,kopietz,kopietzbook,metzner,houghton}.
The dimensional reduction is a purely geometrical procedure of integrating out compact dimensions that emerge in the semiclassical/infrared limit of large distances, $r \gg \lambda_F$, and large times, $\tau \gg 1/E_F$, here $\lambda_F$ and $E_F$ are the Fermi wavelength and the Fermi energy, respectively.
We apply the dimensional reduction to the whole Luttinger-Ward (LW) functional \cite{ward,baym,chitra,kotliar} describing an isotropic spin-degenerate uniform DDEG with $D > 1$ (this procedure is trivial in the $D = 1$ case).

The dimensional reduction procedure is based on the observation that the fermionic correlations exhibit a 1D character at large distances $r \gg \lambda_F$.
This fact, first pointed out in Ref.~\cite{luther}, rests entirely on the existence of the FS: a surface of constant energy, the Fermi energy $E_F$, in the momentum space that separates particle and hole continua at zero temperature $T = 0$.
Here we consider isotropic spin-degenerate DDEG, so the FS is a sphere of radius $k_F$, $k_F = 2 \pi /\lambda_F$ is the Fermi momentum.
A spherical FS results in the equivalence of all points on the FS due to the same character of quantum fluctuations in the vicinity of each point on the FS.
Such an equivalence, thus, identifies the reduced $1 + 1$-dimensional phase space $(\omega, q)$ that is orthogonal to the FS, where $\omega \ll E_F$ is the electron frequency, and $q = p - k _F \ll k_F$ with $p$ being the electron	momentum. 
The spatial dimensions that are orthogonal to this $1 + 1$-dimensional space are compactified for the large-distance fermion correlations with $r \gg \lambda_F$ which results in the 1D-like long-distance asymptotics of the electron Green's function, see Refs.~\cite{miserev22,lounis,miserev21}.
The purpose of dimensional reduction is to integrate out pure geometric effects of these compactified dimensions with the angular measure $\delta \theta \sim	\sqrt{\lambda_F/r} \ll 1$ per each of the $(D - 1)$ compact dimensions at the level of the whole LW functional.

We previously applied the dimensional reduction to a perturbative treatment of non-analytic corrections to the free energy of an interacting DDEG with arbitrary momentum-dependent spin splitting~\cite{miserev22}
and found full agreement of our general result with previously known special cases~\cite{maslov,zak1,zak2}.
This supports the validity of our approach which we extend far beyond the perturbation theory in this paper.
We also stress that the dimensional reduction automatically simplifies all calculations because many degrees of freedom are integrated out universally \cite{miserev22}.
Earlier in Ref.~\cite{miserev21}, we also applied the dimensional reduction to a special case of the resonant exchange scattering in a DDEG within a self-consistent Born approximation where we neglected both forward scattering and the interaction vertex corrections.
In this paper we apply the dimensional reduction to the whole LW functional with arbitrary interaction in the semiclassical/infrared limit.
We stress that this procedure is asymptotically exact in this limit.
The analytically derived and parameter-free semiclassical/infrared limit of the LW functional may be then further exploited for numerical approximations such as DMRG~\cite{schollwock} and GW~\cite{holm,romaniello,houcke}, as well as for various analytic approximations~\cite{miserev22,miserev21}.

Importantly, in the limit when the FLCT is valid \cite{kopietz,neumayr,kopper}, our approach agrees with multidimensional bosonization results \cite{frohlich,marchetti,schwiete,aleiner,efetov,pepin,meier,castellani,kopietz,kopietzbook,metzner,houghton}: the random phase approximation (RPA) then becomes asymptotically exact within the semiclassical/infrared limit.
In this paper we consider a general setting that includes the spectral curvature and the backscattering interaction, such that the FLCT is no longer exact.
We find that skeleton diagrams containing multiple fermion loops are strongly divergent in $D > 1$.
We show that this divergence is sufficient to make the backscattering interaction relevant in $D > 1$ and the spectral curvature relevant in $D > 2$.
The multi-loop skeleton diagrams also represent the non-collinear scattering contribution that is missing in the multidimensional bosonization approach \cite{frohlich,marchetti,schwiete,aleiner,efetov,pepin,meier,castellani,kopietz,kopietzbook,metzner,houghton}.
Such diagrams, which are also poorly studied, might be important for strong correlation effects in interacting DDEGs.

This paper is organized as follows. In Sec.~\ref{sec:LW} we introduce the LW functional of the interacting DDEG. 
The general semiclassical/infrared asymptotic limit of the fermion Green's function and self-energy, as well as the dressed interaction and the polarization operator are derived in Sec.~\ref{sec:asympt}.
The general structure of the dimension-reduced LW functional is presented in Sec.~\ref{sec:DRP}.
The dimensional reduction of all skeleton diagrams containing a single fermion loop is performed in Sec.~\ref{sec:firstorder}, Sec.~\ref{sec:Phiforward}, Sec.~\ref{sec:back}.
The infrared-divergent multi-loop skeleton diagrams representing the non-collinear scattering contributions are considered in Sec.~\ref{sec:noncol}, where general diagrammatic rules for the dimension-reduced LW functional are formulated.
We compare our theory with predictions of the multidimensional bosonization in Sec.~\ref{sec:multi}, where we also demonstrate the relevance of the backscattering and the spectral curvature in higher dimensions.
Conclusions are given in Sec.~\ref{sec:conc}.
Technical details are outlined in Appendices.

\section{Luttinger-Ward formalism}
\label{sec:LW}
	
In order to describe interacting uniform DDEG, we employ the LW formalism~\cite{ward}, also known as the Baym-Kadanoff formalism~\cite{baym}, within the double Legendre transform formulation~\cite{chitra,kotliar}:
\begin{eqnarray}
	&& \mathcal{A}[G, \Sigma, V, \Pi] = - {\rm Tr}\ln \left(G_0^{-1} - \Sigma\right)
	- {\rm Tr}\left\{\Sigma G\right\} \nonumber \\
	&& + \frac{1}{2} \left[{\rm Tr}\left\{\Pi V\right\} + {\rm Tr}\ln\left(V_0^{-1}
	- \Pi\right)\right] + \Phi[G, V] , \label{A}
\end{eqnarray}
where $\mathcal{A}[G, \Sigma, V, \Pi]$ is the LW functional that depends on four bi-local fields $G$, $\Sigma$, $V$, $\Pi$; $G_0$ and $V_0$ are the bare electron Green's function and the bare interaction, respectively; ${\rm Tr}$ stands for the trace over all spin, time and space indices.
The functional $\Phi[G, V]$ is represented as an infinite sum of the two-particle-irreducible, also known as skeleton, diagrams: cutting any two fermion or any two interaction lines must not disconnect a skeleton diagram.
Skeleton diagrams obey standard diagrammatic rules with the following numerical prefactor:
\begin{eqnarray}
&& \frac{(-1)^{n + 1 + F}}{2 n} N_\Sigma, \label{prefactor}
\end{eqnarray}
where $n$ is the number of interaction lines $V$, $F$  the number of fermion loops, and $N_\Sigma$ corresponds to the number of topologically inequivalent graphs derived from the skeleton diagram by cutting a single fermion line.
We note that $N_\Sigma$ is a divisor of $2 n$, the integer $2 n /N_\Sigma$ is called the symmetry factor of a skeleton diagram.
The exact electron Green's function $G$, the self-energy $\Sigma$, the dressed interaction $V$, and the polarization operator $\Pi$ correspond to the saddle-point solutions of the LW functional given by Eq.~(\ref{A}):
\begin{eqnarray}
	&& \frac{\delta \mathcal{A}}{\delta \Sigma} = 0 \,\, \Longleftrightarrow \,\, G
	= \left(G_0^{-1} - \Sigma\right)^{-1} , \label{Gdyson} \\
	&& \frac{\delta \mathcal{A}}{\delta \Pi} = 0 \,\, \Longleftrightarrow \,\, V =
	\left(V_0^{-1} - \Pi\right)^{-1} , \label{Vdyson} \\
	&& \frac{\delta \mathcal{A}}{\delta G} = 0 \,\, \Longleftrightarrow \,\, \Sigma
	= \frac{\delta \Phi[G, V]}{\delta G} , \label{sigmaexact} \\
	&& \frac{\delta \mathcal{A}}{\delta V} = 0 \,\, \Longleftrightarrow \,\, \Pi = -
	2 \frac{\delta \Phi[G, V]}{\delta V} . \label{piexact}
\end{eqnarray}
In this paper we do not introduce separate notations for the saddle-point solutions and the bi-local fields due to clear context: 
the bi-local fields are used within the LW functional, while the saddle-point solutions correspond to the physical	correlation functions that satisfy Eqs.~(\ref{Gdyson})--(\ref{piexact}).
It is clear from Eq.~(\ref{sigmaexact}) that $N_\Sigma$ in Eq.~(\ref{prefactor}) is the number of topologically inequivalent dressed self-energy diagrams generated by a given skeleton diagram of $\Phi[G, V]$.
The functional derivative in Eq.~(\ref{sigmaexact}) generates each self-energy diagram precisely $2 n /N_\Sigma$ times which cancels the symmetry factor in Eq.~(\ref{prefactor}).
Similarly, one can check that Eq.~(\ref{piexact}) generates all diagrams for the polarization operator with correct prefactors.

In general, the full set of saddle-point solutions satisfying 	Eqs.~(\ref{Gdyson})--(\ref{piexact}) contains spurious unphysical 	solutions~\cite{kozik,vucicevic} due to the strong non-linearity of the saddle-point equations.
It has been argued in Refs.~\cite{eder,stan,lindsey} that the spurious solutions can be removed completely by demanding correct analytic properties of the physical Green's function and the dielectric function:
\begin{eqnarray}
	&& G(z, \bm p) = \int\limits_{-\infty}^\infty \frac{\rho_e(\omega, \bm p) \,
		d\omega}{z - \omega} , \hspace{5pt} \rho_e(\omega, \bm p) \ge 0, \label{pos} \\
	&& \frac{V(z, \bm q)}{V_0(z, \bm q)} - 1 = \int\limits_0^\infty
	\frac{\rho_V(\omega, \bm q)}{z^2 - \omega^2} d \left(\omega^2\right) , \hspace{5pt}
	\rho_V(\omega, \bm q) \ge 0,  \label{neg}
\end{eqnarray}
where $\rho_e(\omega, \bm p) \ge 0$, $\rho_V(\omega, \bm q) \ge 0$ are 	positive-definite electron and interaction spectral functions, respectively, $\bm p$ is the electron momentum, $\bm q$ is the interaction momentum, $z$ is a complex frequency, ${\rm Im}(z) \ne 0$.
The Matsubara representation corresponds to $z = i \omega_n$, $\omega_n$ is a fermionic (bosonic) Matsubara frequency in context of the electron Green's function (interaction).
Equation~(\ref{neg}) follows from the Kramers-Kronig relation for the dielectric function~\cite{dolgov}.

In this work we consider a spherical spin-degenerate FS of radius $k_F$, the Fermi momentum.	
According to Ref.~\cite{miserev22}, the results of this paper can be straightforwardly generalized to the case of a non-spherical FS with arbitrary (yet, small enough) momentum-dependent spin splitting.
An important condition here is that each spin-split component of the FS is a smooth manifold with non-zero Gauss curvature at each point, see Ref.~\cite{miserev22}.
In case if the FS contains points of zero Gauss curvature, an additional analysis is required due to anomalously large contributions of these points to the long-range asymptotics of the electron Green's function at particular ``resonant'' directions,  see, e.g., Ref.~\cite{lounis}.
The general results of our work are applicable to interacting DDEGs with  \textit{regular} FS, i.e., any FS (not only spheres) of strictly positive (or strictly negative) Gauss curvature.

We set throughout the reduced Planck constant and the Boltzmann constant to one,	$\hbar = k_B = 1$.

\section{Long-distance and low-energy asymptotics of $G$, $\Sigma$, $V$ and $\Pi$}
\label{sec:asympt}
	
The correlation functions in this paper are expressed in space-time representation rather than momentum-frequency representation.
Here we concentrate on the semiclassical/infrared limit that is commonly driven by strong correlation effects: $r \gg \lambda_F$ and $\tau \gg 1/E_F$, where $\lambda_F = 2 \pi /k_F$ is the Fermi wavelength, $E_F$ is the Fermi energy.

Any nontrivial infrared physics is characterized by the low-energy singularities of correlation functions.
A singularity here is used in a broad sense and it does not necessarily imply the divergence, it can be any kind of discontinuity, e.g., a branch-cut non-analyticity.
The singularities of interacting DDEG are naturally associated with the FS, a manifold in the $D$-dimensional momentum space separating occupied and empty electron states.
The correlation effects are expected to be especially dramatic near the FS where the electron occupation can fluctuate strongly even at zero temperature $T = 0$.
In this paper we assume existence of the FS as a manifold corresponding to the leading singularities of the electron Green's function and self-energy~\cite{senthil,esterlis}.
For simplicity, we also assume that the FS is a $D$-dimensional sphere of radius $k_F$.
Here, we stress that the self-energy singularities emerge even within the Fermi liquid ground state, and that those singularities are responsible for various non-analytic responses \cite{schwiete,aleiner,efetov,pepin,meier,maslov,zak1,zak2,miserev22}.

The singularities near the FS result in the following leading contribution to the semiclassical/infrared asymptotics of the electron Green's function, e.g., see Refs.~\cite{miserev22,lounis,miserev21}: 
\begin{eqnarray}
	&& G(\tau, \bm r) \approx  \frac{e^{i k_F r - i \vartheta}}{\left(\lambda_F 		r\right)^{\frac{D-1}{2}}} g (\tau, r) + \frac{e^{-i k_F r + i		\vartheta}}{\left(\lambda_F r\right)^{\frac{D-1}{2}}} g(\tau, -r) ,
	\label{Gasympt} \\
	&& g(\tau, x) \equiv T  \sum\limits_{\omega_n}\int\limits_{-\infty}^\infty \frac{dq}{2 \pi} e^{i q x - i \omega_n \tau} G(i \omega_n, q) , \label{greduced} \\
	&&  \vartheta = \frac{\pi}{4}(D - 1) , \label{vartheta}
\end{eqnarray}
where $\lambda_F = 2 \pi / k_F$ is the Fermi wavelength,  $D > 1$ is the spatial dimension, $G(i \omega_n, q)$  the electron Green's function in  frequency-momentum 	representation with $\omega_n = \pi T (2 n + 1)$, $n$ integer, being the fermionic Matsubara frequency, $T$ the temperature, and $q = p - k_F$ the distance from the momentum $\bm p$ to the FS. 
The function $g(\tau, x)$, being a 1D Fourier transform of $G(i\omega_n, q)$, represents the effective 1D dual of the original $D$-dimensional Green's function $G(\tau, \bm r)$. 
Importantly, the two terms in Eq.~(\ref{Gasympt}) originate from small vicinities of two points on the spherical FS with the outward normal being collinear to the coordinate vector $\bm r$, see Ref.~\cite{miserev22}.
Each of the $(D - 1)$ dimensions that are tangential to the FS at these points are effectively compactified within the small angular measure $\delta \theta \sim \sqrt{\lambda_F/r} \ll 1$ at $r \gg \lambda_F$, which results in the power-law prefactor in Eq.~(\ref{Gasympt}).
The phase factor $e^{\pm i \vartheta}$ in Eq.~(\ref{Gasympt}) represents the semiclassical phase $\vartheta$ contributing $\pi/4$ per compactified tangential dimension.
A similar asymptotics has been derived in Ref.~\cite{miserev22} for any non-spherical FS with strictly positive (or strictly negative) Gauss curvature, which allows for a straightforward extension of the results of this paper for all such regular FS.

At this point it is useful to introduce the \textit{chiral index} distinguishing between ``left'' and ``right'' components of the 1D dimension-reduced Green's functions, by the following rule:
\begin{eqnarray}
	&& g_R(\tau, x) = g(\tau, x), \,\,\,\, g_L (\tau, x) = g(\tau, -x) . \label{glr}
\end{eqnarray}
So far, the chiral index just allows for more compact representation of Eq.~(\ref{Gasympt}):
\begin{eqnarray}
	&& G(\tau, \bm r) \approx  \sum\limits_{\nu = \pm 1}\frac{e^{i \nu \left(k_F r - \vartheta\right)}}{\left(\lambda_F r\right)^{\frac{D-1}{2}}} g_\nu (\tau, r) ,
	\label{Gasympt2}
\end{eqnarray}
where $\nu = +1$ ($\nu = -1$) corresponds to $\nu = R$ ($\nu = L$) chirality.
In case $D = 1$, Eq.~(\ref{Gasympt2}) coincides with the expansion of 1D Green's function over the left and right movers, e.g., see Ref.~\cite{giamarchi}.
According to the origin of two terms in Eq.~(\ref{Gasympt}) discussed earlier, the chiral index in $D > 1$ spatial dimensions just counts all points on the FS with the outward normals that are collinear to $\bm r$. 
For any regular FS there are exactly two such points at any direction of $\bm r$, see Ref.~\cite{miserev22}.

Exactly the same logic can be applied to the self-energy $\Sigma(\tau, \bm r)$ whose leading long-range asymptotics also originates from the FS singularity, i.e., it also exhibits the asymptotic decomposition of the form of Eq.~(\ref{Gasympt2}): 
\begin{eqnarray}
	&& \Sigma(\tau, \bm r) \approx \sum\limits_{\nu = \pm 1} \frac{e^{i \nu \left(k_F r - \vartheta \right)}}{\left(\lambda_F r\right)^{\frac{D-1}{2}}} s_\nu(\tau, r),
	\label{Sigmaasympt} \\
	&& s(\tau, x) \equiv T \sum\limits_{\omega_n}\int\limits_{-\infty}^\infty \frac{dq}{2 \pi} e^{i q x - i \omega_n \tau} \Sigma(i \omega_n, q) ,
	\label{sigma1d} \\
	&& s_R(\tau, x) = s(\tau, x), \,\,\,\, s_L(\tau, x) = s(\tau, -x) , \label{slr}
\end{eqnarray}
where $\vartheta$ is the semiclassical phase defined in Eq.~(\ref{vartheta}), $s(\tau, x)$ is the 1D Fourier transform of the exact self-energy	$\Sigma(i \omega_n, q)$, with, again, $\omega_n$ being  the fermionic Matsubara frequency and $q = p - k_F$, the distance from the momentum $\bm p$ to the FS.
The chiral components $s_\nu(\tau, x)$ of the effective 1D self-energy are defined similar to the chiral components $g_\nu(\tau, x)$ of the effective 1D Green's function, see Eq.~(\ref{glr}).
According to Eqs.~(\ref{glr}), (\ref{slr}), the 1D Green's function and the 1D self-energy then satisfy the following identities:
\begin{eqnarray}
	&& g_{-\nu}(\tau, -x) = g_{\nu}(\tau, x), \hspace{5pt} s_{-\nu}(\tau, -x) = s_\nu (\tau, x) . \label{gschiral}
\end{eqnarray}

Singularities of the dressed interaction $V(\tau, q)$ originate from the forward 	scattering, $V^{q\sim 0}(\tau, q)$, associated with long-range electron-electron interactions and collective plasmonic effects, 
and from the backscattering, $V^{2 k_F}(\tau, \overline{q})$, which characterizes self-consistent resonant $2 k_F$ scattering:
\begin{eqnarray}
	&& V(\tau, q) \approx V^{q \sim 0}(\tau, q)  + V^{2 k_F}(\tau, \overline{q}) , \label{harmexp}
\end{eqnarray}
where $\overline{q} = q - 2 k_F \ll k_F$, $\tau \gg 1/E_F$ is the imaginary time.
Equation~(\ref{harmexp}) represents the harmonic expansion of the dressed interaction in momentum space, the regular contributions are neglected.
The singularities originating from the higher order harmonics, $4k_F$, $6k_F$, etc., in Eq.~(\ref{harmexp}) require high-energy virtual transitions with the excitation energy $\sim E_F$ and, therefore, are suppressed.
For example, the higher order harmonics in 1D are subleading for weak enough interaction \cite{giamarchi,matveev}. 
In this paper we omit the higher-order harmonics for simplicity only, the dimensional reduction that we develop in this paper allows one to incorporate such effects straightforwardly.

Taking the $D$-dimensional Fourier transform of Eq.~(\ref{harmexp}), we find the spatial asymptotics of the dressed interaction:
\begin{eqnarray}
	&& \hspace{-18pt}  V(\tau, r) = V_1 (\tau, r) + \sum\limits_{\sigma = \pm 1} e^{2 i \sigma \left(k_F r - \vartheta\right)} V_2(\tau, \sigma r), \label{Vdec} \\
	&& \hspace{-18pt} V_1(\tau, r) = \int \frac{d \bm q}{(2 \pi)^D} \, e^{i \bm q \cdot \bm r} V^{q \sim 0}(\tau, q) , \label{forw} \\
	&& \hspace{-18pt} V_2(\tau, x) = \left[\frac{2}{\lambda_F |x|}\right]^{\frac{D - 1}{2}} \!\! e^{i \vartheta \, {\rm sgn}(x)} \! \int\limits_{-\infty}^\infty \! \frac{d\overline{q}}{2 \pi} e^{i \overline{q} x} V^{2 k_F}(\tau, \overline{q}) , \label{back}
\end{eqnarray}
where $\vartheta$ is the semiclassical phase given by Eq.~(\ref{vartheta}), ${\rm sgn}(x)$ returns the sign of $x$, $V_{1}(\tau, r)$, $V_2(\tau, \pm r)$ are slowly varying functions, changing on a scale much larger than $\lambda_F$ at $r \gg \lambda_F$.
Equation~(\ref{forw}) represents the $D$-dimensional Fourier transform of the forward-scattering singularity.
The integration over $\overline{q} = q - 2 k_F$ in Eq.~(\ref{back}) is extended to the whole real line $\mathbb{R}$ due to fast convergence on the scale $\overline{q} \sim 1/r \ll k_F$.
Notice that the asymptotics of the backscattering term in Eq.~(\ref{Vdec}) takes the form that is similar to the Green's function and the self-energy asymptotic expansions, see Eqs.~(\ref{Gasympt2}), (\ref{Sigmaasympt}).
Indeed, the backscattering singularity of the dressed interaction is located on the $(D - 1)$-dimensional sphere of radius $2 k_F$, i.e., we could just use Eq.~(\ref{Gasympt2}) with rescaling $k_F \to 2 k_F$, $\lambda_F \to \lambda_F /2$ and corresponding 1D Fourier transform of $V^{2k_F}(\tau, \overline{q})$ instead of $g(\tau, x)$.
From now on, it is convenient to extend the domain of $V_1(\tau, r)$ to the whole real line $\mathbb{R}$ via the following symmetric extension:
\begin{eqnarray}
	&& V_1(\tau, x) = V_1(\tau, |x|) . \label{V1sym}
\end{eqnarray}

A similar harmonic expansion can be applied to the polarization operator $\Pi$ in the semiclassical/infrared limit $r \gg \lambda_F$, $\tau \gg 1/ E_F$:
\begin{eqnarray}
	&& \hspace{-15pt} \Pi(\tau, q) \approx \Pi^{q \sim 0}(\tau, q) + \Pi^{2 k_F}(\tau, \overline{q}) , \label{Piharm} \\
	&& \hspace{-15pt} \Pi(\tau, r) \approx \frac{P_1 (\tau, r)}{\left(\lambda_F r\right)^{D - 1}}  + \sum\limits_{\sigma = \pm 1} \frac{e^{2 i \sigma \left(k_F r - \vartheta\right)}}{\left(\lambda_F r\right)^{D - 1}} P_2(\tau, \sigma r), \label{Pidec} \\
	&& \hspace{-15pt} P_1(\tau, r) = (\lambda_F r)^{D - 1} \int \frac{d \bm q}{(2 \pi)^D} \, e^{i \bm q \cdot \bm r} \Pi^{q \sim 0} (\tau, q) , \label{landau} \\
	&& \hspace{-15pt} P_2(\tau, x) = \left|2 \lambda_F  x\right|^{\frac{D - 1}{2}} \! e^{i \vartheta \, {\rm sgn}(x)} \! \int\limits_{-\infty}^\infty \frac{d\overline{q}}{2 \pi} e^{i \overline{q} x} \Pi^{2 k_F}(\tau, \overline{q}) . \label{Kohn}
\end{eqnarray}
The power-law prefactor in Eq.~(\ref{Pidec}) is introduced for convenience, the functions $P_{1,2}(\tau, x)$ vary substantially only on a scale much larger than $\lambda_F$ at $ |x| \gg \lambda_F$.

\begin{figure}[t]
	\centering
	\includegraphics[width=0.39\columnwidth]{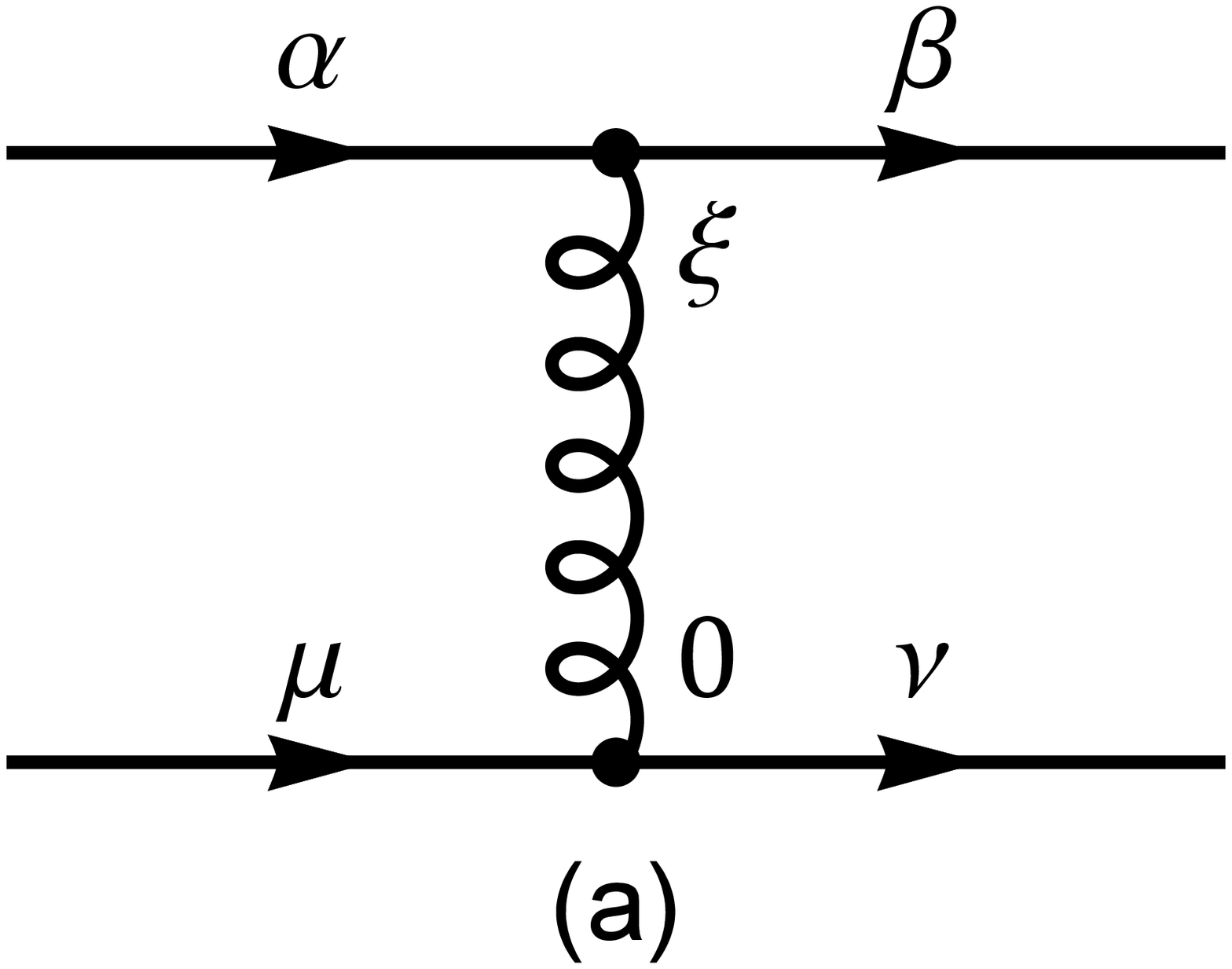}
	\hspace{3mm}
	\includegraphics[width=0.55\columnwidth]{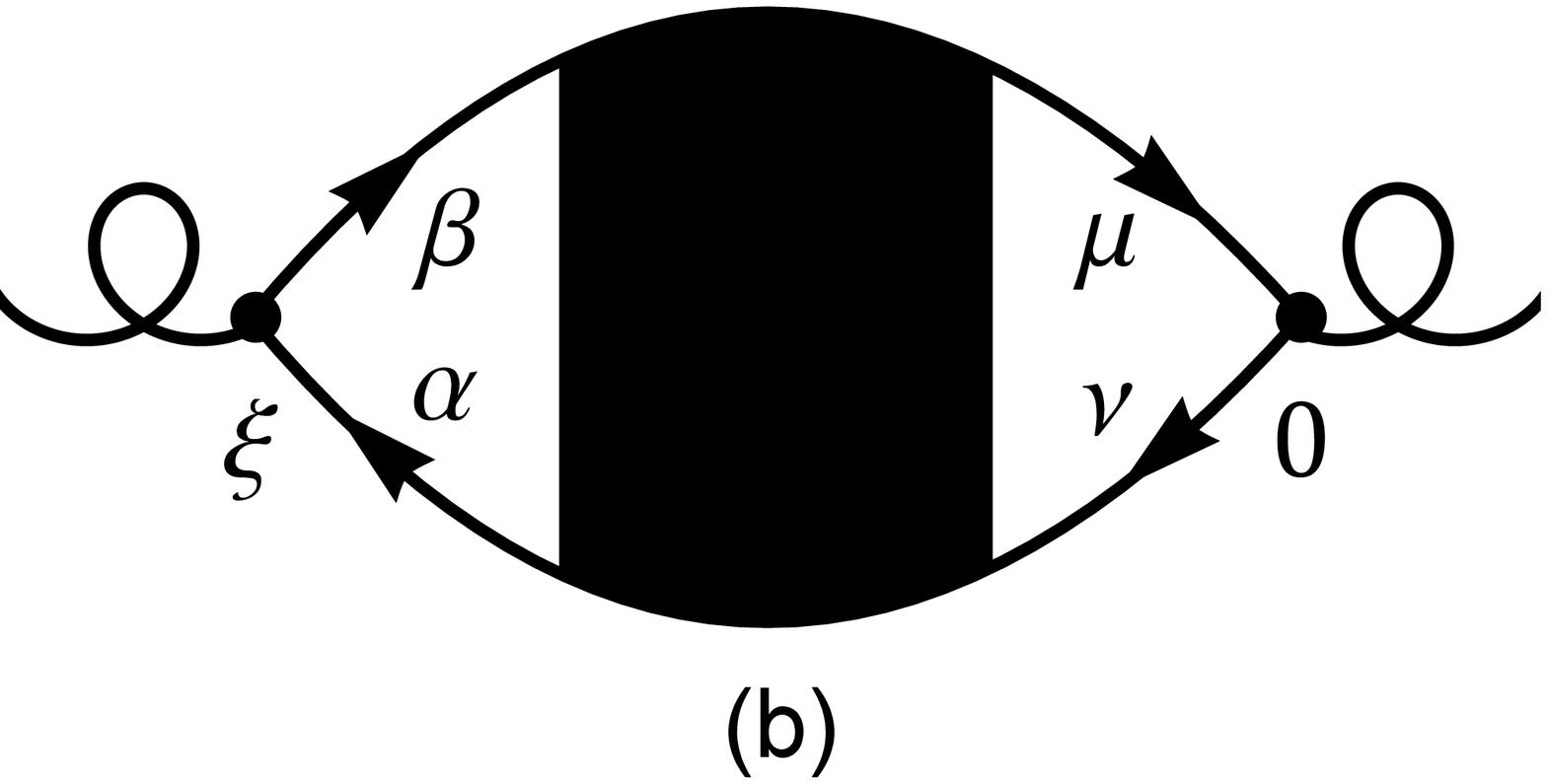}
	\caption{ 
		Chiral matrix elements of: (a) dressed 1D interaction 			$v_{\mu\nu}^{\alpha\beta}(\xi)$,
		(b) dressed 1D polarization operator $\mathcal{P}_{\mu\nu}^{\alpha\beta}(\xi)$, see Eqs.~(\ref{vchiral}), (\ref{Pchiral}), where 	$\xi = (\tau, x)$ and $\alpha, \beta, \mu, \nu \in \{L, R\}$ are the chiral indices.
		}
	\label{fig:chiral}
\end{figure}

Finally, we introduce the chiral indexing convention for the 1D interaction matrix elements as well as for the 1D polarization operator via the following identifications, see Fig.~\ref{fig:chiral}(a),(b):
\begin{eqnarray}
	&& \hspace{-10pt} v^{\nu\nu}_{\mu\mu}(\tau, x) = V_1(\tau, x), \hspace{5pt} v^{\nu -\nu}_{-\nu \nu} (\tau, x) = V_2 (\tau, \nu x), \label{vchiral} \\
	&& \hspace{-10pt} \sum\limits_{\mu, \nu}\mathcal{P}_{\mu\mu}^{\nu\nu}(\tau, x) = P_1(\tau, x) , \hspace{5pt} \mathcal{P}^{-\nu \nu}_{\nu -\nu} (\tau, x) = P_2(\tau, \nu x) , \label{Pchiral}
\end{eqnarray}
where  $\mu$ and $\nu$ are fixed everywhere except in the expression for $P_1(\tau, x)$.
Here, $P_1(\tau, x)$ is represented by the sum of four matrix elements for convenience: in such a representation the chiral indexing acquires its full pseudospin properties.
We emphasize that only the combinations of the chiral components of the effective 1D interaction and 1D polarization operator given in Eqs.~(\ref{vchiral}), (\ref{Pchiral}) are physical.
All other chiral matrix elements that are not shown in Eqs.~(\ref{vchiral}), (\ref{Pchiral}), necessarily vanish in any translation-invariant system due to the momentum conservation: according to our convention, two different chiral fermion species carry momenta $k_F \bm n$ and $-k_F \bm n$ along a certain direction $\bm n$, so the momentum conservation in the DDEG necessarily imposes the conservation of total chiral index in the dimension-reduced effective theory.
Emergence of other chiral components of the effective 1D interaction and the 1D polarization operator that break the total chiral index conservation, corresponds to the momentum conservation in the original DDEG only modulo $2 k_F$, which is equivalent to the emergence of a $2 k_F$ density wave order.
This constitutes the equivalence between the spontaneous translational symmetry breaking of the interacting DDEG and the spontaneous chiral symmetry breaking of the corresponding dimension-reduced low-energy theory that we construct in this work.
A proper discussion of the symmetry breaking requires the Hartree (or condensate) term in Eq.~(\ref{A}), see Ref.~\cite{kotliar}.
We delegate the discussion of ordered states to our future study.

\section{General structure of dimension-reduced LW functional}
\label{sec:DRP}

Substituting the asymptotic expansions of $G$, $\Sigma$, $V$, and $\Pi$, see 	Eqs.~(\ref{Gasympt2}),~(\ref{Sigmaasympt}),~(\ref{Vdec}), and~(\ref{Pidec}), into the LW functional $\mathcal{A}[G, \Sigma, V, \Pi]$, see Eq.~(\ref{A}), we find a new functional, $\tilde \mathcal{A}[g, s, v, \mathcal{P}]$, whose saddle point corresponds to the ground state solutions for $g$, $s$, $v$ and $\mathcal{P}$ defined in Sec.~\ref{sec:asympt}:
\begin{eqnarray}
	&& \tilde \mathcal{A}[g, s, v, \mathcal{P}] = \frac{1}{C_D} \mathcal{A}\left[G[g], \Sigma[s], V[v], \Pi[\mathcal{P}]\right] , \label{tildeA} \\
	&& C_D = \frac{\pi^{\frac{D}{2}}}{\lambda_F^{D-1}
		\Gamma\left(D/2\right)} = \frac{A_{D - 1}}{2 \lambda_F^{D - 1}}, \label{CD}
\end{eqnarray}
where the constant factor $C_D$ is introduced for convenience, 
$\Gamma(x)$ stands for the Euler gamma function, and $A_{D - 1}$ is the surface area of the  $(D - 1)$-dimensional unit sphere, see Eq.~(\ref{area}).
The goal of this work is to simplify Eq.~(\ref{tildeA}).

The first three terms of Eq.~(\ref{A}) yield similar 1D contributions to Eq.~(\ref{tildeA}), see Appendix~\ref{app:A} for details:
\begin{eqnarray}
	&& -\frac{{\rm Tr}\ln \left(G_0^{-1}[g_0] - \Sigma[s]\right)}{C_D} = -{\rm Sp} \ln \left(g_0^{-1} - s\right) , \label{fermred}
	\\
	&& -\frac{{\rm Tr} \left\{\Sigma[s] \, G[g] \right\}}{C_D} = - {\rm Sp} \left\{s \, g\right\}, \label{sgred} \\
	&& \frac{1}{2 C_D} {\rm Tr}\left\{\Pi[\mathcal{P}] \, V[v] \right\} =
	\frac{1}{2} {\rm Sp} \left\{v \, \mathcal{P} \right\} , \label{a3red}
\end{eqnarray}
where ${\rm Sp}$ stands for the 1D trace that includes the integration over the imaginary time $\tau$ and the effective single space coordinate $x$, and also the summation over chiral and spin indices.
Here $G_0[g_0]$ implies the same asymptotic expansion as for $G[g]$, see Eq.~(\ref{Gasympt2}).
The effective dimension-reduced LW functional $\tilde \mathcal{A}[g, s, v, \mathcal{P}]$ can then be represented in the following form:
\begin{eqnarray}
&& \tilde \mathcal{A}[g, s, v, \mathcal{P}] = - {\rm Sp} \ln \left(g_0^{-1} - s\right) - {\rm Sp} \left\{s g\right\} \nonumber \\
&& \hspace{40pt}  + \frac{1}{2} \left[{\rm Sp} \left\{\mathcal{P} v \right\} + \tilde \mathcal{A}\left[\mathcal{P}\right]\right] + \tilde \Phi[g, v] , \label{Atilde} \\
&& \tilde \mathcal{A}\left[\mathcal{P}\right] \equiv \frac{1}{C_D} {\rm Tr} \ln \left(V_0^{-1} - \Pi\left[\mathcal{P}\right]\right) , \label{a4} \\
&& \tilde \Phi[g, v] \equiv \frac{\Phi[G[g], V[v]]}{C_D} , \label{Phitilde}
\end{eqnarray}
where $\tilde \mathcal{A}\left[\mathcal{P}\right]$ describes the polarization effects in the dimension-reduced theory,
$\tilde \Phi[g, v]$ is represented by the dimension-reduced skeleton diagrams that are derived in the subsequent sections.

The saddle-point equations of $\tilde \mathcal{A}[g, s, v, \mathcal{P}]$ allow us to find the 1D duals of the $D$-dimensional correlation functions:
\begin{eqnarray}
	&& \frac{\delta \tilde \mathcal{A}[g, s, v, \mathcal{P}]}{\delta s} = 0 \,\, \Longleftrightarrow \,\, g =
	\left(g_0^{-1} - s\right)^{-1} , \label{gdyson} \\
	&& \frac{\delta \tilde \mathcal{A}[g, s, v, \mathcal{P}]}{\delta g} = 0 \,\, \Longleftrightarrow \,\, s =
	\frac{\delta \tilde \Phi[g, v]}{\delta g} , \label{sexact} \\
	&& \frac{\delta \tilde \mathcal{A}[g, s, v, \mathcal{P}]}{\delta v} = 0 \,\, \Longleftrightarrow \,\,
	\mathcal{P} = - 2 \frac{\delta \tilde \Phi[g, v]}{\delta v} ,
	\label{Pexact} \\
	&& \frac{\delta \tilde \mathcal{A}[g, s, v, \mathcal{P}]}{\delta \mathcal{P}} = 0 \,\, \Longleftrightarrow \,\, v = - \frac{\delta \tilde \mathcal{A}[\mathcal{P}]}{\delta \mathcal{P}} . \label{vpdyson}
\end{eqnarray}
Here we emphasize that Eqs.~(\ref{gdyson})--(\ref{vpdyson}) can be obtained directly from Eqs.~(\ref{Gdyson})--(\ref{piexact}).
However, the dimensional reduction of the LW functional itself allows us to simplify the derivations significantly.
From Eq.~(\ref{gdyson}) we see that the relation between $g$ and $s$ is still given via the standard Dyson equation of the form of Eq.~(\ref{Gdyson}).
The functional $\tilde \Phi[g, v]$ plays the role of the generating functional for the 1D self-energy, $s$, and the 1D polarization operator, $\mathcal{P}$.
In this context, $g$, $s$, and $\mathcal{P}$ can still be represented as a sum of  Feynman diagrams with fully dressed interaction lines.
However, the relation between $v$ and $\mathcal{P}$ can no longer be described via a corresponding diagrammatic series which is clearly seen from the following fact:
\begin{eqnarray}
	&& v \ne \left(v_0^{-1} - \mathcal{P}\right)^{-1} . \label{noQFT}
\end{eqnarray}
This can be directly verified using Eqs.~(\ref{a4}), (\ref{vpdyson}) with the relation between $\Pi$ and $\mathcal{P}$ given in Sec.~\ref{sec:asympt}.
The violation of Wick's theorem of this kind was reported in Ref.~\cite{wen1990} in the context of weakly coupled 2D and 3D arrays of 1D Luttinger liquids.

In practice, it may be convenient to use Eqs.~(\ref{gdyson})--(\ref{Pexact}) that can still be represented in diagrammatic form with all interaction lines dressed.
However, instead of Eq.~(\ref{vpdyson}), one may use the original Eq.~(\ref{Vdyson}) with the identifications $V[v]$ and $\Pi[\mathcal{P}]$ introduced in Sec.~\ref{sec:asympt}.
In the following sections, we derive the effective dimension-reduced representation of $\tilde \Phi[g, v]$.

\begin{figure}[t]
	\centering
	\includegraphics[width=0.5\columnwidth]{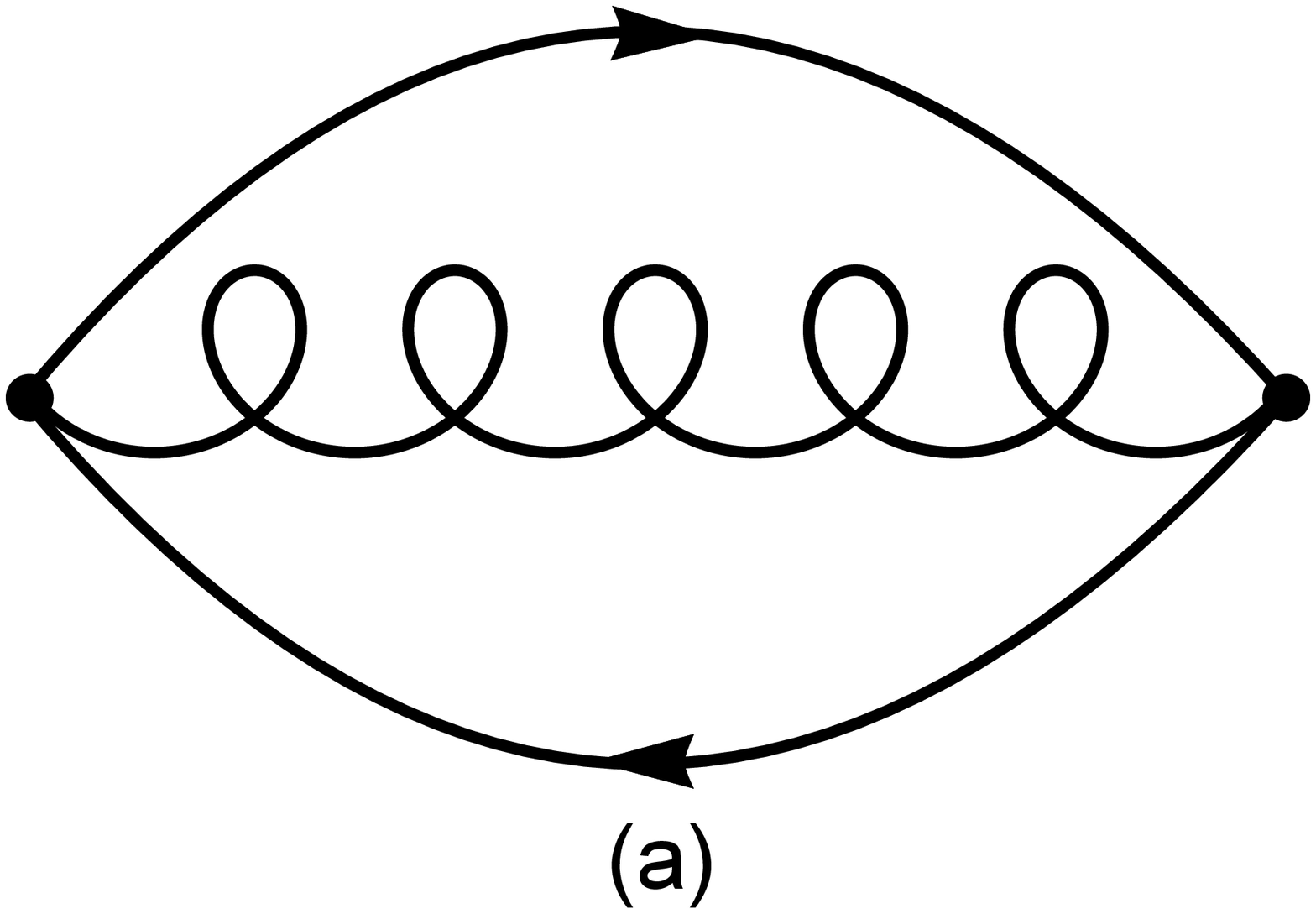}\\
	\includegraphics[width=0.41\columnwidth]{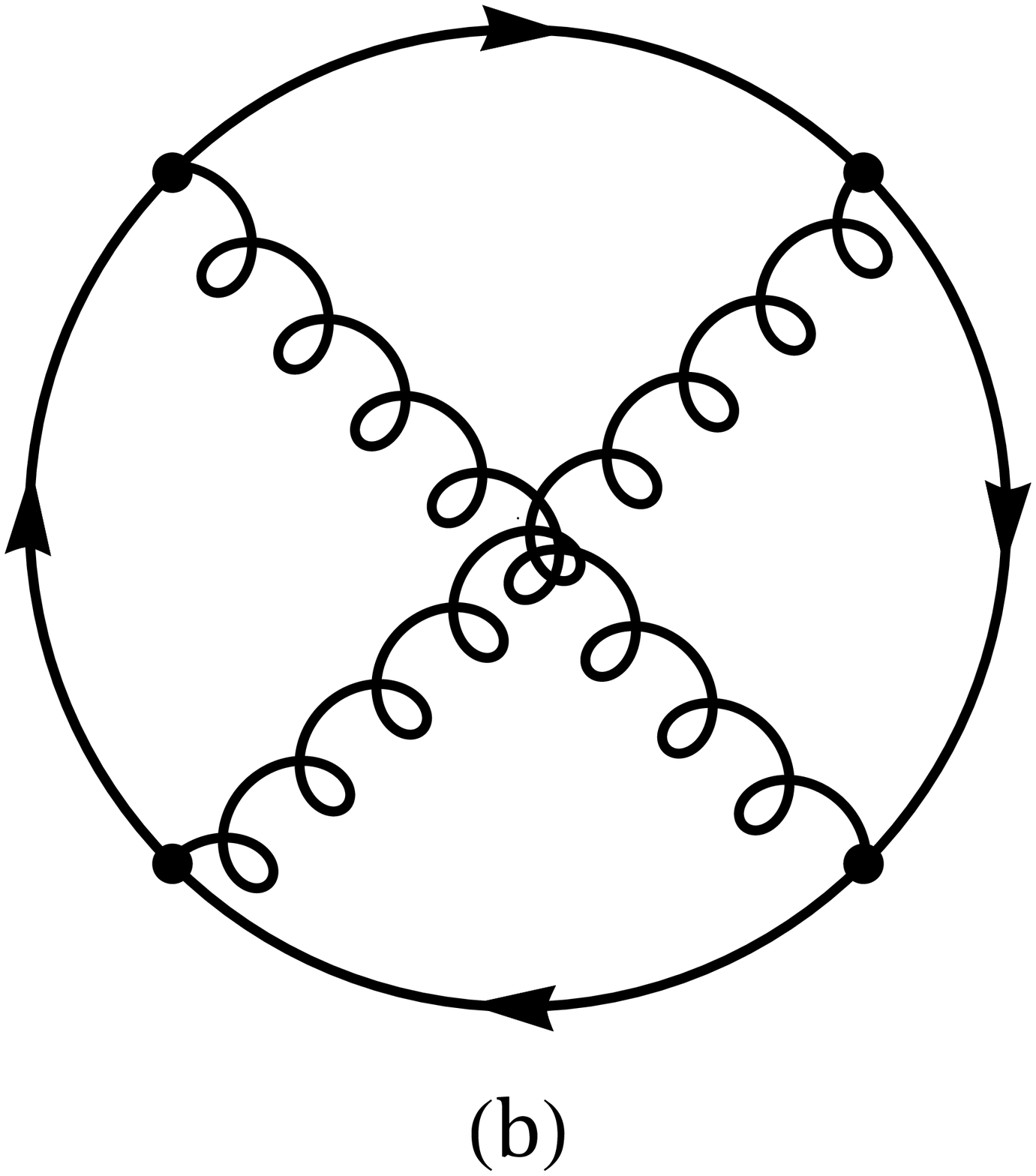} 
	\hspace{3mm}
	\includegraphics[width=0.49\columnwidth]{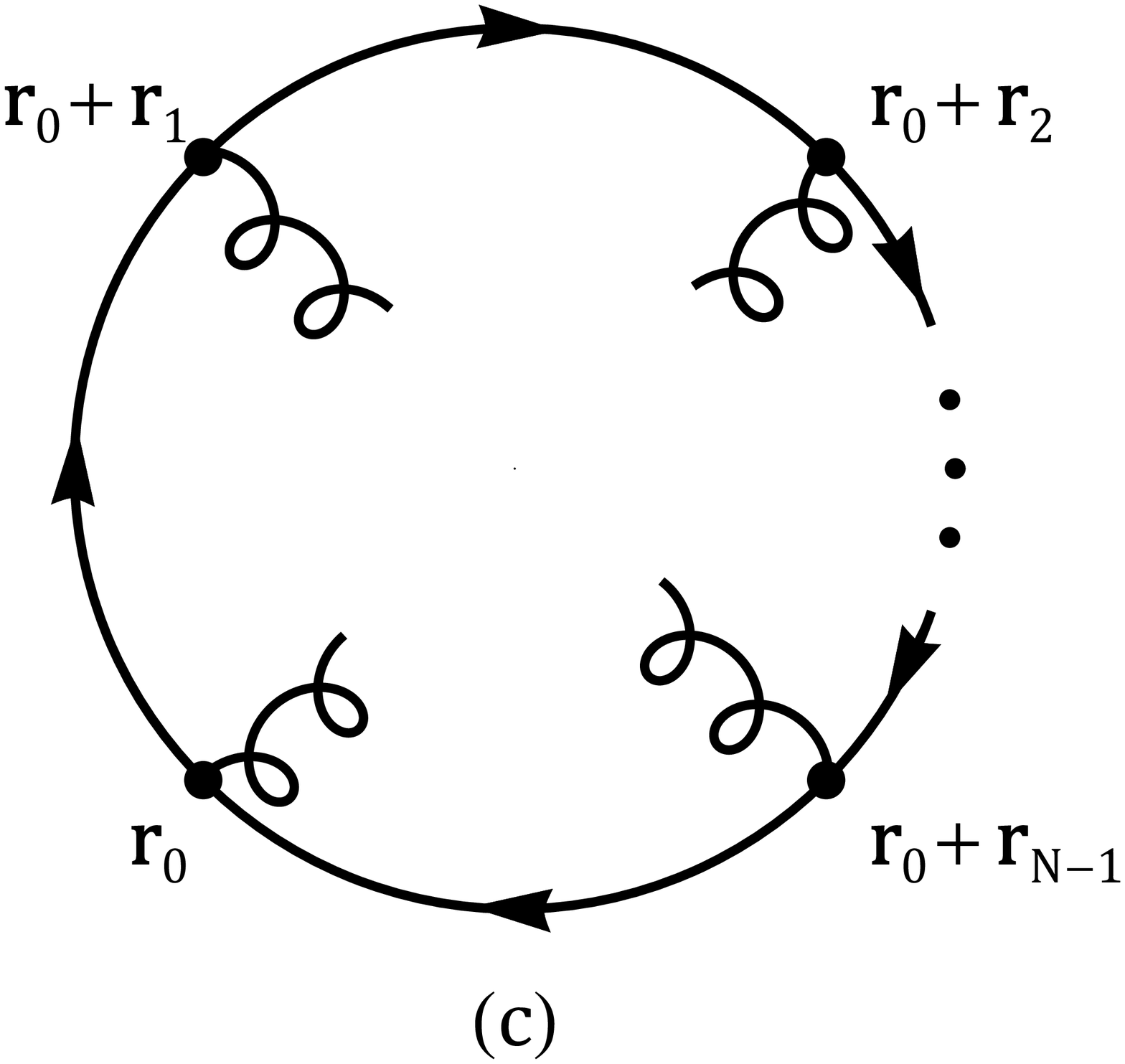}
	\caption{ 
		Skeleton diagrams with a single fermion loop.
		(a) The first-order skeleton diagram with the $2$-vertex fermion loop.
		(b) The only second-order skeleton diagram contains the fermion loop with $N = 4$ vertices.
		(c) Arbitrary skeleton diagram with a single fermion loop and $N$ interaction vertices; here $\bm r_1$, ..., $\bm r_{N - 1}$ are the relative $D$-dimensional spatial coordinates of the loop, $\bm r_0$ is the absolute loop coordinate.
		Here, $N$ must be even as each interaction (wavy) line must connect two separate vertices on the loop.
		The directed solid lines correspond to $G$.
	}
	\label{fig:diagab}
\end{figure}

\section{First-order skeleton diagram}
\label{sec:firstorder}
	
The only skeleton diagram in $\Phi[G, V]$ containing a closed fermion loop with $N = 2$ interaction vertices is the first-order skeleton diagram, see Fig.~\ref{fig:diagab}(a):
\begin{eqnarray}
	&& \Phi_1[G, V] = -\frac{1}{2}\int dz \, V(z) {\rm tr} \left\{G(z) G(-z)\right\}
	, \label{Phi1}
\end{eqnarray}
where $z = (\tau, \bm r)$, ${\rm tr}$ stands for the spin trace.
The truncation of $\Phi[G, V]$ by this diagram constitutes the GW approximation \cite{holm,romaniello,houcke}. 
Substituting the long-distance asymptotics of the Green's function, Eq.~(\ref{Gasympt2}), and the harmonic decomposition of the dressed interaction, Eq.~(\ref{Vdec}), into Eq.~(\ref{Phi1}), we find:
\begin{eqnarray}
	&& \hspace{-20pt} \frac{\Phi_1[G[g], V[v]]}{C_D} = - \int d\tau \int\limits_0^\infty dr
	\sum\limits_{\nu_1, \nu_2} e^{i (k_F r - \vartheta)(\nu_1 + \nu_2)} \nonumber \\
	&& \hspace{10pt} \times  {\rm tr} \left\{g_{\nu_1}(\tau, r) g_{\nu_2}(-\tau, r)\right\}  \nonumber\\
	&& \hspace{10pt} \times \left[V_1(\tau, r) + \sum\limits_{\sigma = \pm 1} e^{2 i \sigma \left(k_F r - \vartheta\right)} V_2(\tau, \sigma r)\right] , \label{phi1}
\end{eqnarray}
where $\nu_1, \nu_2$ are the chiral indices, $C_D$ is given by Eq.~(\ref{CD}), the trivial angular integration has already been performed.
We note that the factor $r^{D - 1}$ in the $D$-dimensional integration measure $d \bm r = A_{D - 1} r^{D - 1} \, dr$, $A_{D - 1}$ is given by Eq.~(\ref{area}), cancels with the power-law factor $1/r^{D - 1}$ coming from the asymptotics of two Green's functions in Eq.~(\ref{Phi1}).
The infrared physics at large scale, $r \gg \lambda_F$, comes from the sector where the fast oscillatory phase in Eq.~(\ref{phi1}) is compensated.
This condition is satisfied at $\nu_2 = - \nu_1$ for the forward-scattering contribution and at $\nu_2 = \nu_1 = -\sigma$ for the backscattering term:
\begin{eqnarray}
	&& \frac{\Phi_1[G[g], V[v]]}{C_D} = - \int d\tau \int\limits_0^\infty dr \, \nonumber \\
	&& \left[V_1(\tau, r) \sum\limits_\nu {\rm tr} \left\{g_\nu(\tau, r) g_\nu
	(-\tau, -r) \right\}\right. \nonumber \\
	&& + \left. \sum\limits_{\nu} V_2(\tau, -\nu r) \, {\rm tr} \left\{g_\nu (\tau, r) g_{-\nu} (-\tau, -r) \right\} \right] , \label{phi12}
\end{eqnarray}
where we relabeled $\nu_1 \to \nu$ and used Eq.~(\ref{gschiral}).
Using that the transformation $\nu \to - \nu$ with $r \to - r$ does not change the expression in square brackets in Eq.~(\ref{phi12}), we can extend the integration over $r$ to the real line $\mathbb{R}$.
Using further the chiral indexing, see Eqs.~(\ref{vchiral}), (\ref{Pchiral}), we find that Eq.~(\ref{phi12}) can be represented in its pure 1D form:	
\begin{eqnarray}
	&& \tilde \Phi_1[g, v] \equiv \frac{\Phi_1[G[g], V[v]]}{C_D} \nonumber \\
	&& \hspace{35pt} = -\frac{1}{2} \sum\limits_{\mu, \nu} \int d\xi \, v^{\mu\nu}_{\nu\mu} (\xi) {\rm tr}
	\left\{ g_\nu(\xi) g_\mu(-\xi) \right\} , \label{Phi11d}
\end{eqnarray}
where $\xi \equiv (\tau, x)$, $x \in (-\infty, \infty)$, $\tau \in (0, 1/T)$, $T$ is the temperature, and $\mu, \nu \in \{L, R\}$ are the chiral indices.
It is clear that $\tilde \Phi_1[g, v]$ represents the same skeleton diagram as in Fig.~\ref{fig:diagab}(a) with the natural identification $G \to g$ and $V \to v$.

\section{Skeleton diagrams with a single fermion loop: forward scattering}
\label{sec:Phiforward}
	
\begin{figure}[t]
	\centering
	\includegraphics[width=0.9\columnwidth]{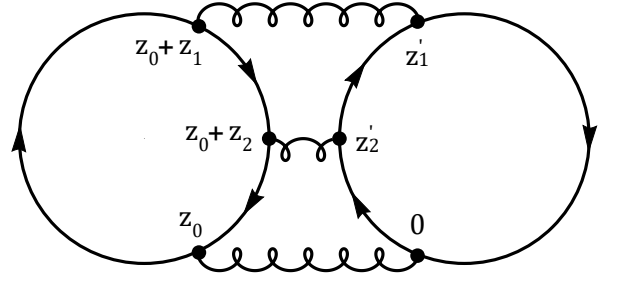}
	\caption{ 
		The third-order skeleton diagram with two $3$-vertex fermion loops.
		The vertex coordinates, $z = (\tau, \bm r)$, are labeled in accord with Eq.~(\ref{Phi3}).
		Here, $z_{1,2}$ and $z_{1,2}'$ are the relative coordinates of the unprimed and primed loops, $z_0$ is the absolute coordinate of the unprimed loop, the absolute coordinate of the primed loop is set to zero.
	}
	\label{fig:thirdorder}
\end{figure}

In this section we perform the dimensional reduction of skeleton diagrams consisting of a single fermion loop with an arbitrary number $N$ of interaction vertices, see Fig.~\ref{fig:diagab}(c).
All interaction lines in this section represent only the forward scattering $V_1$, see Eq.~(\ref{Vdec}), the backscattering is considered in next section.
As each interaction line must connect two separate vertices, then $N$ must be an even number.
The fermion loops with odd number of vertices are possible if a skeleton diagram contains more than one fermion loop, see, for instance, Fig.~\ref{fig:thirdorder}.
Skeleton diagrams with multiple fermion loops are considered in Sec.~\ref{sec:noncol}.
As the time indices are not involved in the dimensional reduction procedure, we do not indicate them for brevity of expressions.
Due to the translation invariance, all two-point functions depend only on the difference of coordinates.
In case of a single-loop diagram all interaction lines must connect two vertices on the same loop.
Therefore, any single-loop diagram is independent of the \textit{absolute loop coordinate} $\bm r_0$, see Fig.~\ref{fig:diagab}(c).
The integration over $\bm r_0$ yields the volume of $D$-dimensional space.
As  we work here with the effective LW functional per unit volume, we can choose an arbitrary value for $\bm r_0$, usually we set $\bm r_0$ to zero.
Thus, the only nontrivial integrations must be performed over the \textit{relative loop coordinates} $\bm r_1, \dots, \bm r_{N - 1}$, see Fig.~\ref{fig:diagab}(c).

First, let us take the integral over $\bm r_1$:
\begin{eqnarray}
	&& \hspace{-10pt} \Phi_N = \int (\dots) \nonumber \\
	&& \int d \bm r_1 \, V_1(\bm r_1 - \bm r_{1'}) {\rm tr} \left\{G(r_1) G(|\bm r_1 - \bm r_2|) \dots
	\right\} , \label{Phinr1}
\end{eqnarray}
where $\Phi_N$ denotes a skeleton diagram in Fig.~\ref{fig:diagab}(c) that consists of a single fermion loop with $N/2$ interaction lines, ${\rm tr}$ stands for the spin trace taken along the fermion loop.
Only the terms that depend on $\bm r_1$ are highlighted in Eq.~(\ref{Phinr1}).
The coordinate $\bm r_{1'} \ne \bm r_1$ represents another vertex connected by the forward-scattering line with $\bm r_1$.
Here it is only important that $\bm r_{1'}$ is fixed during the integration over $\bm r_1$.
Substituting asymptotics of the electron Green's functions, see Eq.~(\ref{Gasympt2}), into Eq.~(\ref{Phinr1}), we find:
\begin{eqnarray}
	&& \hspace{-20pt} \Phi_N = \int (\dots) \int\limits_0^\infty d r_1 \, r_1^{D - 1} \sum\limits_{\nu_1, \nu_2} \frac{e^{-i \nu_1 \left(k_F r_1 - \vartheta \right)}}{\left|\lambda_F r_1\right|^{\frac{D - 1}{2}}} \nonumber \\
	&& \times \int d \bm n_{12} \,  \frac{e^{i \nu_2\left(k_F |\bm r_1 - \bm r_2| - \vartheta \right)}}{\left(\lambda_F \left|\bm r_1 - \bm r_2\right|\right)^{\frac{D - 1}{2}}} V_1(\bm r_1 - \bm r_{1'}) \nonumber \\
	&& \times {\rm tr}\left\{g_{\nu_1}(- r_1) g_{\nu_2}(|\bm r_1 - \bm r_2|) \dots
	\right\} , \label{Phinslow}
\end{eqnarray}
where we integrate over directions of $\bm r_1$ relative to $\bm r_2$, i.e., $d \bm r_1 = r_1^{D - 1} d r_1 \, d \bm n_{12}$.
Notice that for $G(r_1)$ we used Eq.~(\ref{Gasympt2}) with $\nu = -\nu_1$ and accounted for Eq.~(\ref{gschiral}).
As all functions of $\bm r_1$, except the oscillatory exponentials, are slowly varying functions, we can use Eq.~(\ref{Jasympt}) derived in Appendix~\ref{appC}, in order to evaluate the leading contribution coming from the integral over $\bm n_{12}$:
\begin{eqnarray}
	&& \hspace{-20pt} \Phi_N = \int (\dots) \int\limits_0^\infty d r_1 \, \sum\limits_{\nu_1, \nu_2, \sigma_1} \frac{e^{i \vartheta (\nu_1 - \nu_2(1 - \sigma_1))}}{\left|\lambda_F r_2\right|^{\frac{D - 1}{2}}} \nonumber \\
	&& \times e^{i k_F \left[ \nu_2 |r_1 - \sigma_1 r_2| - \nu_1 r_1\right]} V_1\left(|\sigma_1 r_1 \bm n_2 - \bm r_{1'}|\right) \nonumber \\
	&& \times  {\rm tr}\left\{g_{\nu_1}(-r_1) g_{\nu_2}\left(|r_1 - \sigma_1 r_2|\right) \dots
	\right\} , \label{Phi2slow2}
\end{eqnarray}
where two stationary points correspond to $\bm n_1 = \sigma_1 \bm n_2$, $\sigma_1 = \pm 1$, here $\bm n_1 = \bm r_1/ r_1$ and $\bm n_2 = \bm r_2 / r_2$.
Next, we have to make sure that the fast oscillatory phase factor in Eq.~(\ref{Phi2slow2}) (see the second line) is independent of $r_1$, which can be satisfied if the index $\nu_2$ is chosen as follows:
\begin{eqnarray}
	&& \nu_2 = \nu_1 \, {\rm sgn}\left(r_1 - \sigma_1 r_2\right) , \label{nu2}
\end{eqnarray}
where ${\rm sgn}(x)$ returns the sign of $x$. 
Notice that the following combination of the indices which appears in the constant phase factor in Eq.~(\ref{Phi2slow2}) can be then also simplified:
\begin{eqnarray}
	&& \nu_1 - \nu_2 (1 - \sigma_1) = \sigma_1 \nu_1 , \label{combind}
\end{eqnarray}
where $\nu_2$ satisfies Eq.~(\ref{nu2}), and we used that ${\rm sgn}(r_1 + r_2) = 1$ as $r_1 > 0$ and $r_2 > 0$.
In other words, we just performed the summation over $\nu_2$ under the condition that the phase should be independent of $r_1$:
\begin{eqnarray}
	&& \hspace{0pt} \Phi_N = \int (\dots) \int\limits_0^\infty d r_1 \, \sum\limits_{\nu, \sigma_1} \frac{e^{-i \nu \left(k_F r_2 - \vartheta \right)}}{\left|\lambda_F r_2\right|^{\frac{D - 1}{2}}} V_1\left(|\sigma_1 r_1 \bm n_2 - \bm r_{1'}|\right) \nonumber \\
	&& \hspace{20pt} \times  {\rm tr}\left\{g_{\nu}(- \sigma_1 r_1) g_{\nu}\left(\sigma_1 r_1 - r_2\right) \dots
	\right\}\! , \label{Phi2slow3}
\end{eqnarray}
where we used Eq.~(\ref{gschiral}) and introduced a new index notation $\nu = \sigma_1 \nu_1$ in Eq.~(\ref{Phi2slow3}).
The summation over $\sigma_1$ in Eq.~(\ref{Phi2slow3}) extends the integration over $r_1$ to $\mathbb{R}$:
\begin{eqnarray}
	&& \hspace{-20pt} \Phi_N = \int (\dots) \int\limits_{-\infty}^\infty d x_1 \, \sum\limits_{\nu} \frac{e^{-i \nu \left(k_F r_2 - \vartheta \right)}}{\left|\lambda_F r_2\right|^{\frac{D - 1}{2}}} \nonumber \\
	&& \hspace{-20pt} \times V_1\left(|x_1 \bm n_2 - \bm r_{1'}|\right) {\rm tr}\left\{g_{\nu}(- x_1) g_{\nu}\left(x_1 - r_2\right) \dots
	\right\} . \label{Phi2slow4}
\end{eqnarray}

Next, we integrate over $\bm r_2$, see Fig.~\ref{fig:diagab}(c), so let us then also highlight all terms that depend on $\bm r_2$:
\begin{eqnarray}
	&& \hspace{-20pt} \Phi_N = \int (\dots) \int\limits_{-\infty}^\infty d x_1 \int\limits_0^\infty dr_2 \, r_2^{D - 1} \sum\limits_{\nu} \frac{e^{-i \nu \left(k_F r_2 - \vartheta \right)}}{\left|\lambda_F r_2\right|^{\frac{D - 1}{2}}} \nonumber \\
	&& \hspace{0pt}  \times \int d\bm n_{23} \,  V_1\left(|x_1 \bm n_2 - \bm r_{1'}|\right) V_1\left(\bm r_2 - \bm r_{2'}\right)  \nonumber \\
	&& \hspace{0pt} \times  {\rm tr}\left\{g_{\nu}(- x_1) g_{\nu}\left(x_1 - r_2\right) G(\bm r_2 - \bm r_3)  \dots
	\right\} , \label{Phinr2}
\end{eqnarray}
where $d \bm r_2 = r_2^{D - 1} \, dr_2 \, d \bm n_{23}$ and where we measure directions of $\bm r_2$ with respect to $\bm r_3$.
Here, $\bm r_{2'} \ne \bm r_2$ represents the loop coordinate which is connected with $\bm r_2$ by the forward-scattering line and which remains fixed while we integrate over $\bm r_2$.
Notice that after substituting the asymptotics of $G(\bm r_2 - \bm r_3)$ in Eq.~(\ref{Phinr2}), we restore the structure of Eq.~(\ref{Phinslow}).
Just as the integration over $\bm n_{12}$ in Eq.~(\ref{Phinslow}) resulted in the stationary points with $\bm r_1$ and $\bm r_2$ being collinear, the integration over $\bm n_{23}$ in Eq.~(\ref{Phinr2}) yields the stationary points with collinear $\bm r_2$ and $\bm r_3$.
It is now clear how this process propagates along the loop.
In order to understand how this process terminates, we just have to check what happens at the last vertex with the relative loop coordinate $\bm r_{N - 1}$, see Fig.~\ref{fig:diagab}(c):
\begin{eqnarray}
	&& \hspace{0pt} \Phi_N = \int (\dots) \int\limits_{-\infty}^\infty \prod\limits_{i = 1}^{N - 2} \left(d x_i\right) \int d \bm r_{N-1} \sum\limits_{\nu} \frac{e^{-i \nu \left(k_F r_{N - 1} - \vartheta \right)}}{\left|\lambda_F r_{N - 1}\right|^{\frac{D - 1}{2}}} \nonumber \\
	&& \hspace{20pt}  \times \prod\limits_{(j, j')} \left[V_1\left(|x_j \bm n_{N - 1} - \bm r_{j'}|\right)\right]  \nonumber \\
	&& \hspace{20pt} \times {\rm tr}\left\{g_{\nu}(- x_1) g_{\nu}\left(x_1 - x_2\right) \dots G(r_{N - 1}) 
	\right\} , \label{PhinrN}
\end{eqnarray}
where $\bm n_{N - 1} = \bm r_{N - 1}/ r_{N - 1}$ and $(j, j')$ represents a pair of vertices connected by a forward-scattering line.
Now it is time to simplify the arguments of interactions noticing that all $\bm r_{j'}$ in Eq.~(\ref{PhinrN}) correspond to the stationary points of the angular integrals and, therefore, are all collinear.
As all coordinates, except $\bm r_{N - 1}$, are already integrated out,  we can always choose $\bm r_{j'} = x_{j'} \bm n_{N - 1}$.
Note that this is true even if $\bm r_{j'}$ represents $\bm r_{N - 1}$, as $\bm n_{N - 1} = \bm r_{N - 1}/r_{N - 1}$ with $x_{N - 1} = r_{N - 1}$.
Therefore, we conclude the following:
\begin{eqnarray}
	&& |x_j \bm n_{N - 1} - \bm r_{j'}| = |x_j - x_{j'}|,  \label{xj}
\end{eqnarray}
where the pair of indices $(j, j')$ denotes a pair of vertices connected by a corresponding interaction line.
This last argument also removes all dependencies on $\bm n_{N - 1}$, so the integration over $\bm n_{N - 1}$ is trivial.
Substituting the asymptotics of $G(r_{N - 1})$ in Eq.~(\ref{PhinrN}) and accounting for Eq.~(\ref{xj}), we find,
\begin{eqnarray}
	&& \hspace{0pt} \frac{\Phi_N}{C_D} = \int (\dots) \int\limits_{-\infty}^\infty \prod\limits_{i = 1}^{N - 2} \left(d x_i\right) 2 \int\limits_0^\infty d r_{N-1}  \prod\limits_{(j, j')} \left[V_1\left(|x_{jj'}|\right)\right]  \nonumber \\
	&& \hspace{20pt} \times \sum\limits_{\nu} {\rm tr}\left\{g_{\nu}(- x_1) g_{\nu}\left(x_1 - x_2\right) \dots g_\nu(r_{N - 1}) 
	\right\} , \label{PhiN}
\end{eqnarray}
where $C_D$ is given by Eq.~(\ref{CD}), $x_{jj'} = x_j - x_{j'}$, and $(j,j')$ denotes a pair of vertices that are connected by some interaction line, here also $x_{N - 1} = r_{N - 1}$.
Only the non-oscillatory contribution is taken into account in Eq.~(\ref{PhiN}).
Finally, we extend the integration over $r_{N - 1}$ to the integral over $\mathbb{R}$ by noticing that the expression under the integral in Eq.~(\ref{PhiN}) does not change under the following transformation: $\nu \to - \nu$, $x_i \to - x_i$ for all $i \in \left\{1, 2, \dots, N-2 \right\}$ and $r_{N - 1} \to -r_{N - 1}$.
This finally constitutes the dimensional reduction of an arbitrary skeleton diagram with single fermion loop and forward scattering interaction:
\begin{eqnarray}
	&& \hspace{-15pt} \tilde \Phi_N[g, v]  = \int (\dots) \int\limits_{-\infty}^\infty \prod\limits_{i = 1}^{N - 1} \left(d x_i\right) \prod\limits_{(j, j')} \left[V_1\left(|x_{jj'}|\right)\right]  \nonumber \\
	&& \hspace{15pt} \times \sum\limits_{\nu} {\rm tr}\left\{g_{\nu}(- x_1) g_{\nu}\left(x_1 - x_2\right) \dots g_\nu(x_{N - 1}) 
	\right\} . \label{PhiNtilde}
\end{eqnarray}
The part shown by the dots in Eq.~(\ref{PhiNtilde}) corresponds to the time integrals and the constant coming from the diagrammatic rules of the original $D$-dimensional theory, see Eq.~(\ref{prefactor}).
This means that Eq.~(\ref{PhiNtilde}) represents the same diagram shown in Fig.~\ref{fig:diagab}(c) as the original $\Phi_N$ with the natural relabeling $G \to g$ and $V \to v$.
Here, we proved this statement if $V$ and its corresponding $v$ account for the forward scattering interaction only.
In next section we show that this statement remains true even if the backscattering is included.

\section{Skeleton diagrams with a single fermion loop: including the backscattering}
\label{sec:back}

In this section we perform the dimensional reduction of skeleton diagrams with a single fermion loop containing arbitrary number of forward- and backscattering interaction lines.
Here, we employ an inductive proof via the following procedure: in order to do the dimensional reduction of a skeleton diagram with $n_f$ and $n_b$ forward- and backscattering interaction lines, respectively, we first start from the skeleton diagram with the same topology and with all $n_f + n_b$ lines corresponding to the forward scattering, then we substitute the forward scattering lines by the backscattering ones, one by one, until we get to the desired diagram.

Let us start from a single-loop skeleton diagram with $N \ge 4$ interaction vertices (the only skeleton diagram containing the $2$-vertex loop is the first-order diagram, see Fig.~\ref{fig:diagab}(a), it has been considered separately in Sec.~\ref{sec:firstorder}).
We know that all such diagrams acquire the 1D form, see Sec.~\ref{sec:Phiforward}.
Let us now substitute one of the forward scattering interaction lines by the backscattering interaction, see Eq.~(\ref{Vdec}).
Say, this backscattering line connects the vertices $\bm r_0$ and $\bm r_0 + \bm r_k$, where $\bm r_0$ is the absolute loop coordinate, see Fig.~\ref{fig:diagab}(c).
This corresponds to the following substitution in our diagram:
\begin{eqnarray}
	V_1(r_k) \to \sum\limits_{\sigma = \pm 1} V_2(\sigma r_k) e^{2 i \sigma \left(k_F r_k - \vartheta\right)} . \label{V1to2}
\end{eqnarray}
As the interaction is independent of directions of $\bm r_k$ and all other lines correspond to the forward scattering, then we can actually integrate out all $\bm r_i$, $i \ne k$, precisely the way we did  in Sec.~\ref{sec:Phiforward}, resulting in:
\begin{eqnarray}
	&& \hspace{-15pt} \Phi_N^{(1)} = \int \left(\dots\right) \int\limits_{-\infty}^\infty \prod\limits_{i \ne k} \left(d x_i\right)  \,\prod\limits_{(j, j')}\left[V_1\left(x_{jj'}\right)\right] \int d \bm r_k \nonumber \\
	&& \times \sum\limits_{\nu_1, \nu_2} \frac{e^{i \left(\nu_2 - \nu_1\right) \left(k_F r_k - \vartheta\right)}}{\left|\lambda_F r_k\right|^{D - 1}} 
	\sum\limits_{\sigma = \pm 1} V_2(\sigma r_k) e^{2 i \sigma \left(k_F r_k - \vartheta\right)} \nonumber\\
	&& \times {\rm tr} \left\{ g_{\nu_1}(-x_1) \dots g_{\nu_1}(x_{k - 1} - r_k) \right. \nonumber \\
	&& \hspace{70pt} \left. \times g_{\nu_2}(r_k - x_{k+1}) \dots g_{\nu_2}(x_{N - 1}) \right\} , \label{Phiback1}
\end{eqnarray}
where the superscript $^{(1)}$ just indicates a single backscattering line.
Note that here we slightly modified the logic compared to Sec.~\ref{sec:Phiforward} where we were integrating out along a single path $\bm r_1 \to \bm r_2 \to \dots \to \bm r_{N - 1}$.
Here, we integrate out all $\bm r_i$, $i \ne k$, via two separate paths: one is $\bm r_1 \to \dots \to \bm r_k$ and the other is $\bm r_{N - 1} \to \dots \to \bm r_k$, where the last integration over $\bm r_k$ is the only one that is modified by the substitution Eq.~(\ref{V1to2}).
Note that sums over $\nu_1$ and $\nu_2$ appeared here due to two different paths.
The last step is to ensure that the phase in Eq.~(\ref{Phiback1}) is independent of $r_k$ which is satisfied only if $\nu_2 = -\nu_1$ and $\sigma = \nu_1$.
The angular integration over $\bm n_k$ is then trivial:
\begin{eqnarray}
	&& \hspace{-15pt} \tilde \Phi_N^{(1)} = \frac{\Phi_N^{(1)}}{C_D} = \int \left(\dots\right) \int\limits_{-\infty}^\infty \prod\limits_{i= 1}^{N - 1} \left(d x_i\right)  \,\prod\limits_{(j, j')}\left[V_1\left(x_{jj'}\right)\right] \nonumber \\
	&& \times \sum\limits_{\nu} 
	V_2(\nu x_k) {\rm tr} \left\{ g_{\nu}(-x_1) \dots g_{\nu}(x_{k - 1} - x_k) \right. \nonumber \\
	&& \hspace{60pt} \left. \times g_{-\nu}(x_k - x_{k+1}) \dots g_{-\nu}(x_{N - 1}) \right\} , \label{Phiback2}
\end{eqnarray}
where we also extended the integration over $r_k$ to $\mathbb{R}$. 
Notice that according to Eq.~(\ref{vchiral}) $V_2(\nu x_k) = v^{\nu -\nu}_{-\nu\nu}(x_k)$, which just corresponds to standard convolution of the chiral indices.
Thus, we just proved that the insertion of a backscattering line instead of any forward scattering one does not spoil the result: we still get the 1D skeleton diagram with usual spin and chiral pseudospin convolution rules.

Following the inductive argument, we assume that the 1D structure holds after $n_b - 1$ insertions of the backscattering lines.
We need to prove that the insertion of one more backscattering line does not spoil the dimensional reduction.
By choosing one of the vertices of the substituted interaction being the absolute loop coordinate $\bm r_0$, we can just repeat the same steps as in the paragraph above.
This proves that all skeleton diagrams in $\tilde \Phi[g, v]$ with single fermion loop and with the interaction given by Eq.~(\ref{vchiral}) are represented by the effective 1D skeleton diagrams that follow from the original $D$-dimensional ones via the natural identification $G \to g$ and $V \to v$.

For example, let us consider the second-order skeleton diagram shown in Fig.~\ref{fig:diagab}(b):
\begin{eqnarray}
	&& \Phi_2[G, V] = \frac{1}{4} \int dz_1 dz_2 dz_3 \, V(z_1 - z_3) V(z_2)
	\nonumber \\
	&&\times {\rm tr}\left\{G(-z_1) G(z_1 - z_2) G(z_2-z_3) G(z_3)\right\} ,
	\label{Phi2}
\end{eqnarray}
where $z_i = (\tau_i, \bm r_i)$, $i \in \{1,2,3\}$, and ${\rm tr}$ again stands for the spin trace.
Substituting the Green's function asymptotics, see Eq.~(\ref{Gasympt2}), and the harmonic decomposition of dressed interaction, see Eq.~(\ref{Vdec}), we restore the 1D form of the dimension-reduced second-order skeleton diagram:
\begin{eqnarray}
	&& \hspace{-15pt} \tilde \Phi_2[g, v] \equiv \frac{\Phi_2[G[g], V[v]]}{C_D} \nonumber \\
	&& \hspace{20pt} = \frac{1}{4} \int \prod_{i = 1}^3
	\left(d\xi_i\right) \, v^{\mu\nu}_{\alpha\beta}(\xi_1-\xi_3) v^{\nu
		\alpha}_{\beta \mu}(\xi_2) \nonumber \\
	&& \hspace{20pt} \times {\rm tr}\left\{g_\mu(-\xi_1) g_\nu(\xi_1 - \xi_2)
	g_\alpha (\xi_2 - \xi_3) g_\beta (\xi_3) \right\} , \label{Phi21d}
\end{eqnarray}
where $\xi_i = (\tau_i, x_i)$, $i \in \{1,2,3\}$ and
$\alpha, \beta, \mu, \nu \in \{L, R\}$ are the chiral indices.

\section{Skeleton diagrams with multiple fermion loops: contribution of the non-collinear scattering}
\label{sec:noncol}

\begin{figure}[t]
	\centering
	\includegraphics[width=0.92\columnwidth]{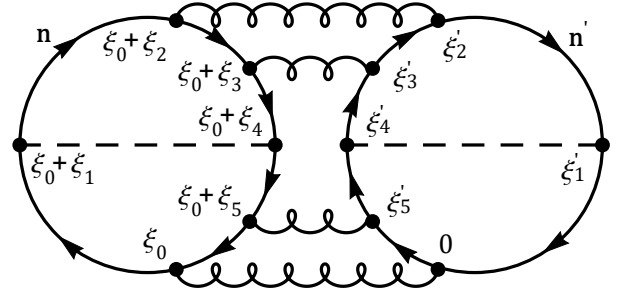}
	\caption{ 
		The sixth-order skeleton diagram with two $6$-vertex fermion loops, both containing the internal backscattering line (dashed line).
		The wavy lines correspond to the forward-scattering interaction here.
		We denoted the relative coordinates $\xi_i = (\tau_i, x_i)$, $\xi_i' = (\tau_i', x_i')$, $i \in \{1, \dots, 5\}$, of the dimension-reduced diagram, see Eq.~(\ref{Phi4}),
		$\xi_0$ is the absolute coordinate of the unprimed loop, the absolute coordinate of the primed loop is set to zero.
		Here, $\bm n$ and $\bm n'$ are the relative loop directions defining the saddle-point values of the $D$-dimensional coordinates: $\bm r_i = \bm n x_i$, $\bm r_i' = \bm n' x_i'$,  $i \in \{1, \dots, 5\}$.
	}
	\label{fig:6order}
\end{figure}

In previous sections we integrated out all relative coordinates on a fermion loop with even number of vertices.
It is clear that the same proofs are also applicable to the fermion loops with odd number of interaction vertices, e.g., see Fig.~\ref{fig:thirdorder}.
The first important difference of the multi-loop skeleton diagrams from the single-loop ones is that some interaction lines might connect two vertices that belong to different fermion loops, see Figs.~\ref{fig:thirdorder}, \ref{fig:6order}.
Within leading order in $\lambda_F/ r$, all such lines must correspond to the forward-scattering interaction: any oscillatory interaction harmonic, $2 k_F$, $4 k_F$, etc., necessarily pins the relative coordinate directions on the connected loops to the same direction, $\pm \bm n$, while there is no such constraint imposed by the forward scattering.
This is especially obvious for the backscattering $2 k_F$ interaction: the backscattering processes are resonant near the FS only if the total momentum of scattered electrons is close to zero, i.e., the backscattering is resonant in the Cooper channel.
There is no such constraint for the forward scattering: momenta of the scattered electrons can be completely uncorrelated as soon as both are near the FS.
The latter represents the non-collinear scattering effect.
Another important difference of the $K$-loop skeleton diagrams from the single-loop ones that are considered in Sec.~\ref{sec:firstorder}, Sec.~\ref{sec:Phiforward}, and Sec.~\ref{sec:back}, comes from the integration over the absolute loop coordinates, see definition of $\bm r_0$ in Fig.~\ref{fig:diagab}(c): 
only one of $K > 1$ absolute loop coordinates can be set to zero due to the translation invariance, say $\bm r_0^{(K)} = 0$; the integration over each of the remaining $K - 1$ absolute loop coordinates, $\bm r_0^{(k)}$, $k \in \{1, \dots, K - 1\}$, contributes the infrared-divergent $D$-dimensional volume factor $\propto |r_0^{(k)}|^{D - 1}$.
This behavior of the multi-loop skeleton diagrams at $D > 1$ is qualitatively different from the genuine 1D case.

For better understanding of the dimensional reduction in the multi-loop case, we show the derivation for the two-loop skeleton diagram in Fig.~\ref{fig:thirdorder}:
\begin{eqnarray}
&& \hspace{-10pt} \Phi_3 = \frac{1}{6} \int dz \, dz' \, \mathcal{L}_3\left(z_1, z_2\right) \mathcal{L}_3\left(z_2', z_1'\right) \nonumber\\
&& \hspace{10pt} \times V_1(z_0) V_1(z_0 + z_1 - z_1') V_1(z_0 + z_2 - z_2') , \label{Phi3}
\end{eqnarray}
where $z = (\tau, \bm r)$, $\tau$ is the imaginary time, $\bm r$ is the $D$-dimensional coordinate, $dz = dz_0 \, dz_1 \, dz_2$, $dz' = dz_1'\, dz_2'$, $V_1(z)$ is the forward-scattering interaction, $\mathcal{L}_3$ denotes the following fermion loop:
\begin{eqnarray}
&& \hspace{-15pt} \mathcal{L}_3(z_1, z_2) = {\rm tr} \left\{ G(-z_1) G(z_1 - z_2) G(z_2)\right\} , \label{L3} 
\end{eqnarray}	
where ${\rm tr}$ stands for the spin trace.
After performing the dimensional reduction over the relative loop coordinates following the lines of reasoning in Sec.~\ref{sec:Phiforward}, we find:
\begin{eqnarray}
&& \hspace{-15pt} \tilde \Phi_3 = \frac{\Phi_3}{C_D} = \frac{C_D}{6} \int d\xi \, d\xi' \, \tilde \mathcal{L}_3\left(\xi_1, \xi_2\right)  \tilde \mathcal{L}_3\left(\xi_2', \xi_1'\right) \nonumber \\
&& \hspace{0pt} \times \int dz_0 \int \frac{d \bm n}{A_{D - 1}} \int \frac{d \bm n'}{A_{D - 1}}  V_1(r_0) \nonumber \\
&& \times  V_1(\bm r_0 + \bm n x_1 - \bm n' x_1') V_1(\bm r_0 + \bm n x_2 - \bm n' x_2') , \label{Phi3reduced} \\
&& \hspace{-15pt} \tilde \mathcal{L}_3(\xi_1, \xi_2) = \sum\limits_{\nu} {\rm tr} \left\{g_\nu (-\xi_1) g_\nu(\xi_1 - \xi_2) g_\nu(\xi_2) \right\} . \label{L3tilde}
\end{eqnarray}
Here, $\xi = (\tau, x)$, $d\xi = d\xi_1 d\xi_2$, $d \xi' = d\xi_1' d\xi_2'$, $C_D$ is given by Eq.~(\ref{CD}), and, again, $A_{D - 1}$ is the surface area of the $(D - 1)$-dimensional unit sphere $S_{D - 1}$.
The time arguments of the forward scattering interactions are not shown explicitly in Eq.~(\ref{Phi3reduced}) for brevity of expressions.
As before, the leading contribution to each loop in Fig.~\ref{fig:thirdorder} comes from the sector where all relative coordinates within each loop are collinear, the general direction for such relative coordinates is denoted here by $\bm n$ and $\bm n'$ for two loops in Fig.~\ref{fig:thirdorder}.
We call such $\bm n$ and $\bm n'$ directions the \textit{relative loop direction} here.
In contrast to Secs.~\ref{sec:Phiforward}, \ref{sec:back}, the angular integrals over the relative loop directions $\bm n$ and $\bm n'$ are no longer trivial because the forward-scattering lines connecting different loops depend on $\bm n$ and $\bm n'$.
We can further simplify Eq.~(\ref{Phi3reduced}) using that $d z_0 = d\tau_0 \, r_0^{D - 1} dr_0 \, d\bm n_0$ and Eq.~(\ref{CD}):
\begin{eqnarray}
&& \hspace{-10pt} \tilde \Phi_3 = \frac{1}{6} \left[\frac{A_{D - 1}}{2}\right]^2 \int d\xi \, d\xi' \, \tilde \mathcal{L}_3\left(\xi_1, \xi_2\right)  \tilde \mathcal{L}_3\left(\xi_2', \xi_1'\right) \nonumber \\
&& \hspace{-10pt} \times \int d\xi_0 \int \frac{d \bm n_0}{A_{D - 1}}\int \frac{d \bm n}{A_{D - 1}} \int \frac{d \bm n'}{A_{D - 1}} \, \left|\frac{x_0}{\lambda_F}\right|^{D - 1}  V_1(x_0) \nonumber \\
&& \hspace{-10pt} \times  V_1(\bm n_0 x_0 + \bm n x_1 - \bm n' x_1') V_1(\bm n_0 x_0 + \bm n x_2 - \bm n' x_2') , \label{Phi3reducedsimp}
\end{eqnarray}
where we also extended the integral over $r_0$ to $\mathbb{R}$ using the change $\bm n_0 \to -\bm n_0$, $\bm n \to - \bm n$, $\bm n' \to - \bm n'$; $d\xi_0 = d\tau_0 dx_0$, $x_0 \in \mathbb{R}$.
The power of $A_{D - 1}/2$ in the first line of Eq.~(\ref{Phi3reducedsimp}) comes from $K - 1$ $C_D$ factors (one $C_D$ factor per loop, minus one comes from the normalization in Eq.~(\ref{Phi3reduced})) and from the normalization of $K - 1$ integrals over the absolute coordinate directions (one absolute coordinate is set to zero, $\bm r_0^{(K)} = 0$), so the overall power is $2 (K - 1)$.
In case of the two-loop diagram in Fig.~\ref{fig:thirdorder}, $K = 2$.

In general, the dimension-reduced diagram with $K$ fermion loops has the following diagrammatic structure:
(i) the order-dependent factor is inherited from the $D$-dimensional representation, see Eq.~(\ref{prefactor});
(ii) multiply it by $[A_{D-1}/2]^{2 (K - 1)}$;
(iii) substitute all loops by their complete 1D analogs (include all internal interaction lines in the definition of such loops);
(iv) set one of the absolute loop coordinates to zero, say $\xi_0^{(K)} = 0$, then integrate over other $K - 1$ absolute coordinates $d\xi_0^{(k)}$, $k \in \{1, \dots, K-1\}$, with the measure factor $|x_0^{(k)}/\lambda_F|^{D - 1}$;
(v) integrate over all absolute coordinate directions, $\bm n_0^{(k)}$, $k \in \{1, \dots, K - 1\}$, and over the relative loop directions, $\bm n$, with the unit measure $d \bm n/A_{D - 1}$.
These simple diagrammatic rules allow us to express arbitrary skeleton diagram of the LW functional in the semiclassical/infrared limit.

Following the diagrammatic rules formulated in the paragraph above, the dimension-reduced skeleton digram shown in Fig.~\ref{fig:6order} takes the following form:
\begin{eqnarray}
&& \tilde \Phi_6 = -\frac{3}{12} \left[\frac{A_{D - 1}}{2}\right]^2 \! \int \! d\xi \, d\xi' \, \tilde \mathcal{L}_6 (\xi_1, \dots, \xi_5) \tilde \mathcal{L}_6 (\xi_5', \dots, \xi_1') \nonumber \\
&& \times \int d\xi_0 \int \frac{d \bm n_0}{A_{D - 1}} \int \frac{d\bm n}{A_{D - 1}} \int \frac{d\bm n'}{A_{D - 1}} \left|\frac{x_0}{\lambda_F}\right|^{D - 1} V_1(x_0) \nonumber \\
&& \times V_1\left(\bm n_0 x_0 + \bm n x_2 - \bm n' x_2'\right) V_1\left(\bm n_0 x_0 + \bm n x_3 - \bm n' x_3'\right) \nonumber \\
&& \times V_1\left(\bm n_0 x_0 + \bm n x_5 - \bm n' x_5'\right) , \label{Phi4}
\end{eqnarray}
where $d\xi = \prod_{i = 1}^5 (d\xi_i)$, $d\xi' = \prod_{i = 1}^5(d\xi_i')$, the order-dependent coefficient $-3/12$ comes from the original $D$-dimensional diagram,
see Eq.~(\ref{prefactor}) with $n = 6$, $F = 2$, and $N_\Sigma = 3$,
the absolute coordinate of the primed loop is set to zero, 
$\bm n_0$ is the direction of the absolute coordinate of the unprimed loop,
$\bm n$ and $\bm n'$ are the relative directions of the primed and unprimed loops,
and the vertex numbering is shown in Fig.~\ref{fig:6order}.
As both loops in Fig.~\ref{fig:6order} are topologically identical, we use a single notation $\tilde \mathcal{L}_6$:
\begin{eqnarray}
&& \hspace{-20pt} \tilde \mathcal{L}_6(\xi_1, \xi_2, \xi_3, \xi_4, \xi_5) = \sum\limits_{\mu, \nu} v^{\mu\nu}_{\nu\mu}(\xi_{14}) \nonumber \\
&& \hspace{-10pt} \times {\rm tr} \left\{g_\mu(-\xi_1) g_\nu(\xi_{12}) g_\nu(\xi_{23}) g_\nu(\xi_{34}) g_{\mu}(\xi_{45}) g_\mu(\xi_5) \right\} , \label{L6}
\end{eqnarray}
where $\xi_{ij} = \xi_i - \xi_j$.
The diagram in Fig.~\ref{fig:6order} corresponds to the backscattering line contribution, $\mu = - \nu$, in Eq.~(\ref{L6}).

In this section we formulated simple diagrammatic-like rules for the dimension-reduced skeleton diagrams.
We note that the non-collinear scattering is explicitly accounted for within our approach, in contrast to the multidimensional bosonization approaches where these effects are omitted.
We emphasize that the dimensional reduction can be applied not only to the LW functional but also to the thermodynamic potential that can be represented by the vacuum bubble diagrams of similar structure.

\section{Multidimensional bosonization: FLCT and beyond}
\label{sec:multi}

In this section we compare our results with predictions of the multidimensional bosonization \cite{haldane,neto,fradkin,delacretaz,frohlich,marchetti,schwiete,aleiner,pepin,efetov,meier,castellani,kopietz,kopietzbook,metzner,houghton}.
We also comment on why the multi-loop diagrams considered in Sec.~\ref{sec:noncol} may be important in higher dimensions $D > 1$ as soon as the spectral curvature and the backscattering are taken into account.

First, let us neglect the spectral curvature and the backscattering.
Under these conditions, the 1D FLCT is exact \cite{dzyalo}.
As all relative coordinates in fermion loops can be reduced to their 1D analogues, see Secs.~\ref{sec:Phiforward}-\ref{sec:noncol}, then the skeleton diagrams containing more than one fermion loop do not contribute to neither $\mathcal{P}$ nor $s$.
We emphasize that the 1D FLCT \cite{dzyalo} is applicable here due to Eqs.~(\ref{gdyson})--(\ref{Pexact}): $g$, $s$, and $\mathcal{P}$ can be expressed in terms of the Feynman diagrams with fully dressed interaction lines but with bare $g_0$.
As a result, only the bare particle-hole bubble, $\mathcal{P}_0$, contributes to $q \sim 0$ component of the polarization operator $\mathcal{P}$:
\begin{eqnarray}
&& \mathcal{P}^{\nu\nu}_{\mu\mu}(\xi) \approx \mathcal{P}_{0\, \mu \mu}^{\nu\nu}(\xi) = \delta_{\mu \nu} {\rm tr} \left\{g_{0\, \nu}(\xi) g_{0\, \nu} (-\xi) \right\} , \label{P0}
\end{eqnarray}
where, again, ${\rm tr}$ stands for the spin trace, $\delta_{ab}$ is the Kronecker index, $g_{0}(\xi)$ is the 1D Fourier transform of $G_0(i \omega_n, q)$, see Eq.~(\ref{greduced}), and $\xi = (\tau, x)$.
Using Eq.~(\ref{Pchiral}), we find the $q \sim 0$ component of the polarization operator:
\begin{eqnarray}
&& \hspace{-15pt} P_1(\tau, x) \nonumber \\
&& = {\rm tr} \left\{g_0(\tau, x) g_0 (-\tau, -x) + g_0(\tau, -x) g_0(-\tau, x) \right\} . \label{P1RPA}
\end{eqnarray}
Substituting Eq.~(\ref{P1RPA}) into Eq.~(\ref{Pidec}), we find that $D$-dimensional polarization operator is also given by the bare particle-hole bubble, $\Pi_0$:
\begin{eqnarray}
&& \Pi(\tau, r) \approx \Pi_0(\tau, r) = {\rm tr} \left\{G_0(\tau, \bm r) G_0(-\tau, - \bm r) \right\} . \label{Pi0}
\end{eqnarray}
Here, the bare electron Green's function $G_0(\tau, \bm r)$ is represented via its asymptotic form, see Eq.~(\ref{Gasympt}), and only $q \sim 0$ component of $\Pi(\tau, r)$ has to be taken into account.
Thus, we indeed confirm that the RPA approximation is asymptotically exact as soon as the spectral curvature and the backscattering are neglected, which agrees with the multidimensional bosonization \cite{haldane,neto,fradkin,delacretaz,frohlich,marchetti,schwiete,aleiner,pepin,efetov,meier,castellani,kopietz,kopietzbook,metzner,houghton}.

Let us now include a finite spectral curvature that is naturally present in the effective mass approximation, and the backscattering that is generated self-consistently.
Under these conditions the FLCT is no longer exact.
These perturbations are known to be irrelevant in 1DEGs,  see, e.g., Refs.~\cite{giamarchi,metzner}.
However, large infrared-divergent measure factors of the multi-loop skeleton diagrams, see Sec.~\ref{sec:noncol}, might be strong enough to make these perturbations relevant in higher dimensions.
The importance of the multi-loop diagrams for the non-analytic corrections in 2DEGs has been pointed out in Ref.~\cite{maslov}, where it has been shown that the diagram in Fig.~\ref{fig:thirdorder} considered in the context of the thermodynamic potential is responsible for the non-collinear scattering contribution to the infrared non-analyticities.
This demonstrates that the multi-loop diagrams are important for the infrared physics of a DDEG with $D > 1$.
In contrast to the multidimensional bosonization approaches, our theory accounts naturally for the non-collinear scattering contribution to the semiclassical/infrared limit of interacting DDEG that goes beyond the FLCT.

In order to quantify the infrared divergence of a $K$-loop skeleton diagram, we introduce the divergence exponent, $\alpha(K)$, which counts additional powers of $|x_0^{(k)}/\lambda_F| \gg 1$, $k \in \{1, \dots, K - 1\}$, coming from the absolute loop coordinates, see Sec.~\ref{sec:noncol}:
\begin{eqnarray}
&& \alpha(K) = \left(K - 1\right) (D - 1) . \label{alpha}
\end{eqnarray}
From this, we conclude that the most infrared-divergent diagrams must contain a large number $K$ of fermion loops,  i.e., those loops must contain a minimal number of vertices.
The multi-loop skeleton diagrams are possible in all orders $N \ge 3$, for instance, a third-order two-loop diagram is shown in Fig.~\ref{fig:thirdorder}.
Note that the $3$-vertex loops are the minimal possible loops in $N^{\rm th}$-order skeleton diagrams with $N \ge 3$ [the $2$-vertex loop is only possible in the first-order skeleton diagram shown in Fig.~\ref{fig:diagab}(a)]. 
Thus, the skeleton diagrams with maximal possible number of  $3$-vertex loops, $K_3(N)$, have the largest divergence exponent $\alpha(K)$, see Eq.~(\ref{alpha}):
\begin{eqnarray}
	&& \hspace{-15pt} K_3(N) = \left\lfloor \frac{2 N}{3} \right\rfloor, \hspace{5pt} \alpha_3(N) = \left[ \left\lfloor \frac{2 N}{3} \right\rfloor - 1\right] \left(D - 1\right) , \label{KN}
\end{eqnarray}
where $\lfloor x \rfloor$ is the floor function, $\alpha_3(N) = \alpha[K_3(N)]$, $N \ge 3$.
The two-particle irreducibility of skeleton diagrams requires all three vertices on any $3$-vertex loop to be external, i.e., all three vertices are connected with other fermion loops, e.g., see Fig.~\ref{fig:thirdorder}.

The divergence exponent $\alpha_3(N)$ of the maximally infrared-divergent diagrams is overestimated.
Indeed, one can show that all odd-vertex fermion loops must vanish in the presence of  particle-hole symmetry which is asymptotically exact in the semiclassical/infrared limit.
The spectral curvature must be taken into account to break the particle-hole symmetry explicitly.
Being a subleading effect, the spectral curvature results in a small $\lambda_F/r$ factor per fermion loop, thus reducing the divergence exponent from $\alpha_3(N)$, see Eq.~(\ref{KN}), to $\alpha_3'(N)$:
\begin{eqnarray}
&& \alpha_3'(N) = \alpha_3(N) - K_3(N) \nonumber \\
&& \hspace{31pt} = \left[ \left\lfloor \frac{2 N}{3} \right\rfloor - 1\right] (D - 2) - 1 . \label{alpham}
\end{eqnarray}
We see that at any $D > 2$ there exists $N_0 \ge 3$ such that $\alpha_3'(N) > 0$ for all $N \ge N_0$.
This means that the spectral curvature is a relevant perturbation at $D > 2$ and thus, the RPA is no longer asymptotically exact. 
We point out that the spectral curvature is qualitatively important for the low-temperature transport properties even in 1DEG where it is formally an irrelevant perturbation \cite{schmidt}.

As we see, the emergent particle-hole symmetry in the semiclassical/infrared limit significantly reduced the divergence exponent of the skeleton diagrams containing maximal number of the $3$-vertex loops.
Instead, we may consider fermion loops with even number of vertices as those are not sensitive to the particle-hole symmetry.
In order to violate the FLCT, we insert one backscattering line within each loop.
As the number of vertices connected with other loops must be strictly greater than two (due to the two-particle irreducibility), the minimal number of vertices is then equal to six.
The smallest such diagram is shown in Fig.~\ref{fig:6order}. 
The divergence exponent, $\alpha_6 (N)$, of skeleton diagrams containing maximal possible number of the $6$-vertex loops is the following:
\begin{eqnarray}
	&& \alpha_6(N) = \left[\left\lfloor \frac{2 N}{6}\right\rfloor - 1 \right] \left(D - 1\right) . \label{alphab}
\end{eqnarray}
As we see, the divergence exponent is strictly positive in any $D > 1$ at $N \ge 6$, emphasizing the relevance of the multi-loop diagrams in higher dimensions.
Exponents $\alpha_3'(N)$ and $\alpha_6(N)$ allow for a new classification of the infrared-divergent diagrams. We believe that this classification may result in new well controlled non-perturbative approaches in strongly correlated electron systems.

In this section we demonstrated that our theory agrees with the multidimensional bosonization results if the FLCT is satisfied.
However, here we also argue that the spectral curvature and the backscattering, both are irrelevant in the 1DEG and both violate the FLCT, may become relevant in higher dimensions due to additional infrared divergence of the multi-loop skeleton diagrams at $D > 1$, see Sec.~\ref{sec:noncol}.
It has been also pointed out in Refs.~\cite{castellani,metzner} that the electron Green's function evaluated self-consistently within the RPA, acquires unphysical singularities in the semiclassical/infrared limit near the single-particle pole line which may also signal that the RPA is not sufficient for an accurate description of the semiclassical/infrared limit of the interacting DDEG at $D > 1$.

\section{Conclusions}
\label{sec:conc}
	
In this work, we have presented a new powerful theoretical tool, the dimensional reduction, that allows for the asymptotically exact treatment of an interacting DDEG, $D > 1$, in the semiclassical/infrared limit where the  effect of interaction is strongest.
Using the LW approach \cite{ward,baym,chitra,kotliar}, we show that the single-loop skeleton diagrams are reduced to effective 1D form.
Together with the FLCT, this is equivalent to the exactness of the RPA in the semiclassical/infrared limit of interacting DDEG, which agrees with conclusions of the multidimensional bosonization \cite{,haldane,fradkin,neto,delacretaz,frohlich,marchetti,schwiete,aleiner,efetov,pepin,meier,castellani,kopietz,kopietzbook,metzner,houghton}.
Skeleton diagrams containing large number of fermion loops represent the non-collinear scattering contribution and are infrared-divergent at $D > 1$.
We show that this divergence makes the spectral curvature relevant at $D > 2$ and the backscattering relevant at $D > 1$, both perturbations explicitly violate the FLCT, and both are irrelevant in 1DEG.
This makes the FLCT unreliable in the semiclassical/infrared limit at $D > 1$.
Our theory still retains simple diagrammatic structure (in terms of irreducible diagrams) which is important for practical calculations, the non-collinear scattering processes that are missing in the multidimensional bosonization approaches, are naturally accounted for here.
Therefore, we believe that the semiclassical/infrared limit of the LW functional of interacting DDEG that we derived in this paper, may step beyond well known predictions of the multidimensional bosonization.
The dimensional reduction technique that is applied here to the LW functional, is quite versatile, it can be straightforwardly generalized for the thermodynamic potential that is represented by similar vacuum bubble diagrams, it can be also applied to perturbation theory corrections, e.g. see Ref.~\cite{miserev22}, or self-consistent approximations, see Ref.~\cite{miserev21}.
The dimensional reduction of interacting DDEG with an arbitrary spin splitting and spontaneously broken symmetries is the subject of our future study.

\section*{Acknowledgments}
We thank Mikhail Pletyukhov for very important stimulating discussions.
We are grateful to Herbert Schoeller for critical reading of the manuscript, valuable comments, and fruitful discussions.
This work was supported by the Georg H. Endress Foundation, the Swiss National Science Foundation (SNSF), and NCCR SPIN.
This project received funding from the European Union's Horizon 2020 research and innovation program (ERC Starting Grant, grant agreement No 757725).\\

	\appendix
	
	\section{Dimensional reduction of the free terms in the LW functional}
	\label{app:A}
	
	Here we derive Eqs.~(\ref{fermred}), (\ref{sgred}), (\ref{a3red}) presented in Sec.~\ref{sec:DRP}.

	In order to simplify the first term in Eq.~(\ref{A}), we use the frequency-momentum representation:
	\begin{eqnarray}
	&& \hspace{-20pt} {\rm Tr}\ln \left(G_0^{-1} -
	\Sigma\right) \nonumber \\
	&& \hspace{-20pt} =  T \sum\limits_{\omega_n} \! \int \! \frac{d \bm p}{(2 \pi)^D} {\rm tr} \left\{ \ln \left[G_0^{-1}(i \omega_n, \bm p) - \Sigma(i \omega_n, \bm p)\right] \right\} , \label{A1}
	\end{eqnarray}
	where ${\rm tr}$ stands for the spin trace.
	Contribution of the infrared sector near the FS comes from $p \approx k_F$ which allows us to simplify the integration measure:
	\begin{eqnarray}
	&& \frac{d\bm p}{(2 \pi)^D} \approx \left[\frac{k_F}{2 \pi}\right]^{D - 1} A_{D - 1} \, \frac{dq}{2 \pi} = 2 C_D \frac{dq}{2 \pi}, \label{dp}\\
	&& A_{D-1} = \frac{2 \pi^{\frac{D}{2}}}{\Gamma\left(D/2\right)} , \label{area} 
	\end{eqnarray}
	where $q = p - k_F \ll k_F$, 
	$A_{D-1}$ is the surface area of the $(D-1)$-dimensional unit sphere $S_{D-1}$,
	$\Gamma(x)$ is the Euler gamma function,
	$C_D$ is given in Eq.~(\ref{CD}).
	Using Eqs.~(\ref{greduced}), (\ref{sigma1d}) as definitions of the effective 1D Green's function and 1D self-energy, we find:
	\begin{eqnarray}
	&& G_0 (i \omega_n, \bm p) = g_0(i \omega_n, q) , \hspace{5pt} \Sigma(i \omega_n, \bm p) = s(i \omega_n, q) , \label{g0s}
	\end{eqnarray}
	where $q = p - k_F \ll k_F$.
	Substituting Eqs.~(\ref{dp}), (\ref{g0s}) back into Eq.~(\ref{A1}), we find the infrared contribution of the first term in Eq.~(\ref{A}):
	\begin{eqnarray}
	&& \hspace{-20pt} \frac{{\rm Tr}\ln \left(G_0^{-1}[g_0] -
	\Sigma[s]\right)}{C_D} \nonumber \\
	&& \hspace{-20pt} = T \sum\limits_{\omega_n}
	\int\limits_{-\infty}^\infty \frac{dq}{2\pi} \, 2 \,  {\rm tr} \left\{\ln
	\left[g_0^{-1}(i \omega_n, q) - s(i \omega_n, q)\right] \right\} ,
	\label{A1red}
	\end{eqnarray}
	where we extended the integration over $q$ to the interval $q \in (-\infty, 	\infty)$.
	Using the chiral indexing introduced in Eqs.~(\ref{glr}), (\ref{slr}) and corresponding chiral symmetry, see Eq.~(\ref{gschiral}), we incorporate the factor of $2$ in Eq.~(\ref{A1red}) into trace over the chiral index:
	\begin{eqnarray}
	&& \hspace{-20pt} \frac{{\rm Tr}\ln \left(G_0^{-1}[g_0] -
		\Sigma[s]\right)}{C_D} = {\rm Sp} \ln \left(g_0^{-1} - s\right) \nonumber \\
	&& \hspace{-20pt} = T \sum\limits_{\omega_n, \nu}
	\int\limits_{-\infty}^\infty \frac{dq}{2\pi} {\rm tr} \left\{\ln \left((g_0^{-1})_{\nu}(i
	\omega_n, q) - s_\nu(i \omega_n, q)\right) \! \right\} \! , \label{A1nu}
	\end{eqnarray}
	where $\nu \in \{L, R\}$ is the chiral index, ${\rm Sp}$ stands for the 1D trace that includes the frequency and 1D momentum summation as well as the spin and chiral traces, see the second line in Eq.~(\ref{A1nu}).

	In order to reduce the dimensionality of the second term in Eq.~(\ref{A}), it is more convenient to use the space-time asymptotics given by Eqs.~(\ref{Gasympt2}), (\ref{Sigmaasympt}):
	\begin{eqnarray}
	&& {\rm Tr}\left\{\Sigma[s] G[g]
	\right\} = \sum\limits_{\nu_i} \int d\tau \int\limits_0^\infty dr \, r^{D
		- 1} A_{D -1} \nonumber \\
	&& \times \frac{e^{i (\nu_1 + \nu_2)(k_F r - \vartheta)}}{\left(\lambda_F
		r\right)^{D - 1}} {\rm tr} \left\{g_{\nu_1}(\tau, r) s_{\nu_2}(-\tau, r)\right\}
	, \label{sg}
	\end{eqnarray}
	where $\nu_{1,2} \in \{L, R\}$ are the chiral indices, ${\rm tr}$ is the spin
	trace.
	Here we used that $d \bm r = A_{D - 1} r^{D - 1} \, dr$, $A_{D-1}$ is given by
	Eq.~(\ref{area}).
	The case $\nu_2 = \nu_1$ corresponds to the integration over fast oscillatory
	terms $e^{\pm 2 i k_F r}$ that are irrelevant to the infrared physics.
	Thus, the only relevant terms in Eq.~(\ref{sg}) correspond to $\nu_2 = -\nu_1$.
	The remaining sum over $\nu_1$ together with the chiral symmetry, see Eq.~(\ref{gschiral}), allows us to extend the integration over $r$ to the real line $\mathbb{R}$, yielding:
	\begin{eqnarray}
	&& \hspace{0pt} \frac{{\rm Tr}\left\{\Sigma[s] G[g]
	\right\}}{C_D}  = {\rm Sp} \left\{s g\right\} \nonumber \\
	&& \hspace{0pt} = \int d\tau
	\int\limits_{-\infty}^\infty dx \, \sum\limits_\nu {\rm tr} \left\{g_\nu(\tau,
	x) s_\nu(-\tau, -x)\right\} , \label{A2}
	\end{eqnarray}
	where ${\rm Sp}$ represents full 1D trace.

	The dimensional reduction procedure of the third term in Eq.~(\ref{A}) is similar to the integration of the second term:
	\begin{eqnarray}
	&& \hspace{-10pt} {\rm Tr} \left\{\Pi V\right\} = \int d\tau \int\limits_0^\infty dr \, r^{D - 1} A_{D - 1} V(\tau, r) \Pi(-\tau, r) , \label{A3a}
	\end{eqnarray}
	where the trivial angular integration is already performed.
	Then, we substitute the harmonic expansions of the interaction and the polarization operator, see Eqs.~(\ref{Vdec}), (\ref{Pidec}):
	\begin{eqnarray}
	&& \frac{{\rm Tr} \left\{\Pi[\mathcal{P}] V[v]\right\}}{C_D} = 2 \int d\tau \int\limits_0^\infty dr \,
	\left[V_1(\tau, r) P_1(-\tau, r) \right. \nonumber \\
	&& \left. + \sum\limits_{\sigma = \pm 1} V_2(\tau, \sigma r) P_2(-\tau, -\sigma r) \right] , \label{A3int}
	\end{eqnarray}
	where we omitted all fast oscillatory terms the same way we did it in Eq.~(\ref{A2}).
	The symmetric extension of $V_1$, see Eq.~(\ref{V1sym}), and the sum over $\sigma$ in Eq.~(\ref{A3int}), allows us to extend the integration over $r \in (0, \infty)$ to the real line $\mathbb{R}$:
	\begin{eqnarray}
	&& \frac{{\rm Tr} \left\{\Pi[\mathcal{P}] V[v]\right\}}{C_D} = \int d\tau \int\limits_{-\infty}^\infty dx \, \left[ V_1(\tau, x) P_1(-\tau, - x) \right. \nonumber \\
	&& \hspace{30pt} \left. + \sum\limits_{\nu = \pm 1} V_2(\tau, \nu x) P_2(-\tau, - \nu x) \right] , \label{A3taux}
	\end{eqnarray}
	where $x \in \mathbb{R}$ is the effective 1D coordinate.
	Finally, we rewrite Eq.~(\ref{A3taux}) in terms of the chiral components of the interaction and the polarization operator introduced in Eqs.~(\ref{vchiral}), (\ref{Pchiral}):
	\begin{eqnarray}
	&&\hspace{-20pt}\frac{{\rm Tr} \left\{\Pi[\mathcal{P}] V[v]\right\}}{C_D} = {\rm Sp} \left\{v \mathcal{P} \right\} \nonumber \\
	&& \hspace{30pt} = \int d\tau
	\int\limits_{-\infty}^\infty dx \, v^{\mu\nu}_{\alpha \beta}(\tau, x)
	\mathcal{P}^{\alpha\beta}_{\mu\nu}(-\tau, - x)  , \label{A31D}
	\end{eqnarray}
	where ${\rm Sp}$ stands for the 1D trace, the second line of Eq.~(\ref{A31D}) also defines the convolution rule of the chiral indices for $v$ and $\mathcal{P}$ that is consistent with the definition, see Eqs.~(\ref{vchiral}), (\ref{Pchiral}) and Fig.~\ref{fig:chiral}(a),(b).

\section{Useful angular integral}
\label{appC}
	
In this appendix we outline the asymptotic behavior of the following integral:
\begin{eqnarray}
	&& J_\nu^Q[f] \equiv \int d \bm n_1 \, e^{i \nu Q |\bm r_1 - \bm r_2|} f(\bm
	r_1, \bm r_2) , \label{J}
\end{eqnarray}
where $\nu = \pm 1$, $Q$ is the large parameter here, the integral is taken over directions $\bm n_1 = \bm r_1/r_1$, $r_1$ and $\bm r_2$ are fixed. 
The function $f(\bm r_1, \bm r_2)$ in Eq.~(\ref{J}) varies slowly on the scale $r_{1,2} \sim 1/Q$ and otherwise, is arbitrary.
The asymptotics of $J^Q_\nu[f]$ can be derived via the stationary phase method.
The extrema of the phase, $Q |\bm r_1 - \bm r_2|$, as a function of $\bm n_{1}$ correspond to $\bm n_1 = \pm \bm n_2$, $\bm n_2 = \bm r_2/r_2$.
The contribution of these two stationary points to the large-$Q$ asymptotics is then the following:
\begin{eqnarray}
	&& \hspace{-13pt} J^Q_\nu[f] \approx \sum\limits_{\sigma = \pm 1} f(\sigma r_1 \bm n_2,
	\bm r_2) e^{i \nu Q |r_1 - \sigma r_2|} j_{\nu}^\sigma(r_1, r_2) ,
	\label{jsta}\\
	&& \hspace{-13pt} j_\nu^\sigma (r_1, r_2) \equiv A_{D - 2} \int\limits_0^\infty
	d\theta \, \theta^{D - 2} \exp\left(i \frac{ \nu \sigma Q r_1 r_2 \theta^2}{2
		|r_1 - \sigma r_2|}\right) ,  \label{jint}
\end{eqnarray}
where we expanded the measure and the phase in a small vicinity of each of two stationary points, $A_{D - 2}$ is the area of a $(D-2)$-dimensional unit sphere, see Eq.~(\ref{area}).
The integral given by Eq.~(\ref{jint}) is reduced to Euler gamma function after the transformation $\theta \to \theta(x)$:
\begin{eqnarray}
	&& \theta(x) = e^{i \frac{\pi}{4} \nu \sigma} \sqrt{\frac{2 |r_1 - \sigma r_2|}{Q r_1 r_2}} \sqrt{x} , \\
	&& j_\nu^\sigma (r_1, r_2) = \left[\frac{2 \pi |r_1 - \sigma r_2|}{Q r_1
		r_2}\right]^{\frac{D - 1}{2}} e^{i \sigma \nu \vartheta} , \label{jsignu}
\end{eqnarray}
where $\vartheta$ is given by Eq.~(\ref{vartheta}).
Substituting Eq.~(\ref{jsignu}) back into Eq.~(\ref{jsta}), we find the asymptotics of $J^Q_\nu[f]$:
\begin{eqnarray}
	&& J^Q_\nu[f] \approx \sum\limits_\sigma f(\sigma r_1 \bm n_2, \bm r_2) e^{i \nu
		\left(Q |r_1 - \sigma r_2| + \sigma \vartheta\right)} \nonumber \\
	&& \hspace{40pt} \times  \left[\frac{2 \pi |r_1 - \sigma r_2|}{Q r_1
		r_2}\right]^{\frac{D - 1}{2}} , \label{Jasympt}
\end{eqnarray}
where $\sigma = \pm 1$, $\bm n_2 = \bm r_2/ r_2$.

\end{document}